\documentclass[3p]{elsarticle}
\usepackage{lineno,hyperref}
\usepackage{amsmath}
\usepackage{booktabs}
\usepackage{float}
\usepackage[normalem]{ulem}
\usepackage{color, soul}
\modulolinenumbers[1]
\usepackage{subfigure}
\usepackage{graphicx}
\usepackage{times}
\usepackage{epsfig}
\usepackage{graphicx}
\usepackage{amsmath}
\usepackage{amssymb}
\usepackage{graphics}
\usepackage{subfigure}
\usepackage{epstopdf}
\usepackage{array}
\usepackage{multirow}
\usepackage{color}
\usepackage{array}
\usepackage{booktabs}
\usepackage{comment}
\usepackage{dcolumn}% Align table columns on decimal point
\usepackage{bm}% bold math
\usepackage{hyperref}% add hypertext capabilities
\usepackage{float}
\usepackage{multirow}
\usepackage{setspace}
\usepackage{amssymb}
\begin{document}

\begin{frontmatter}



\title{Multiscale reconstruction of porous media based on multiple dictionaries learning}


\author[]{Pengcheng Yan}
\ead{yanpengcheng@stu.scu.edu.cn}

\author[]{Qizhi Teng\corref{cor}}
\ead{qzteng@scu.edu.cn}

\author[]{Xiaohai He}
\ead{hxh@scu.edu.cn}

\author[]{Zhenchuan Ma}
\ead{2017851167@qq.com}

\author[]{Ningning Zhang}
\ead{15555325269@163.com}

\cortext[cor]{Corresponding author.}
\address{College of Electronics and Information Engineering, Sichuan University,  Chengdu 610065, China}

\begin{abstract}
Digital modeling of the microstructure is important for studying the physical and transport properties of porous media. Multiscale modeling for porous media can accurately characterize macro-pores and micro-pores in a large-FoV (field of view) high-resolution three-dimensional pore structure model. This paper proposes a multiscale reconstruction algorithm based on multiple dictionaries learning, in which edge patterns and micro-pore patterns from homology high-resolution pore structure are introduced into low-resolution pore structure to build a fine multiscale pore structure model. The qualitative and quantitative comparisons of the experimental results show that the results of multiscale reconstruction are similar to the real high-resolution pore structure in terms of complex pore geometry and pore surface morphology. The geometric, topological and permeability properties of multiscale reconstruction results are almost identical to those of the real high-resolution pore structures. The experiments also demonstrate the proposal algorithm is capable of multiscale reconstruction without regard to the size of the input. This work provides an effective method for fine multiscale modeling of porous media.

\end{abstract}


\end{frontmatter}

%\linenumbers


\section{\label{sec:introduction}Introduction}

\begin{figure*}[h]
	\centering
	\includegraphics[keepaspectratio=True, width=0.45\textwidth]{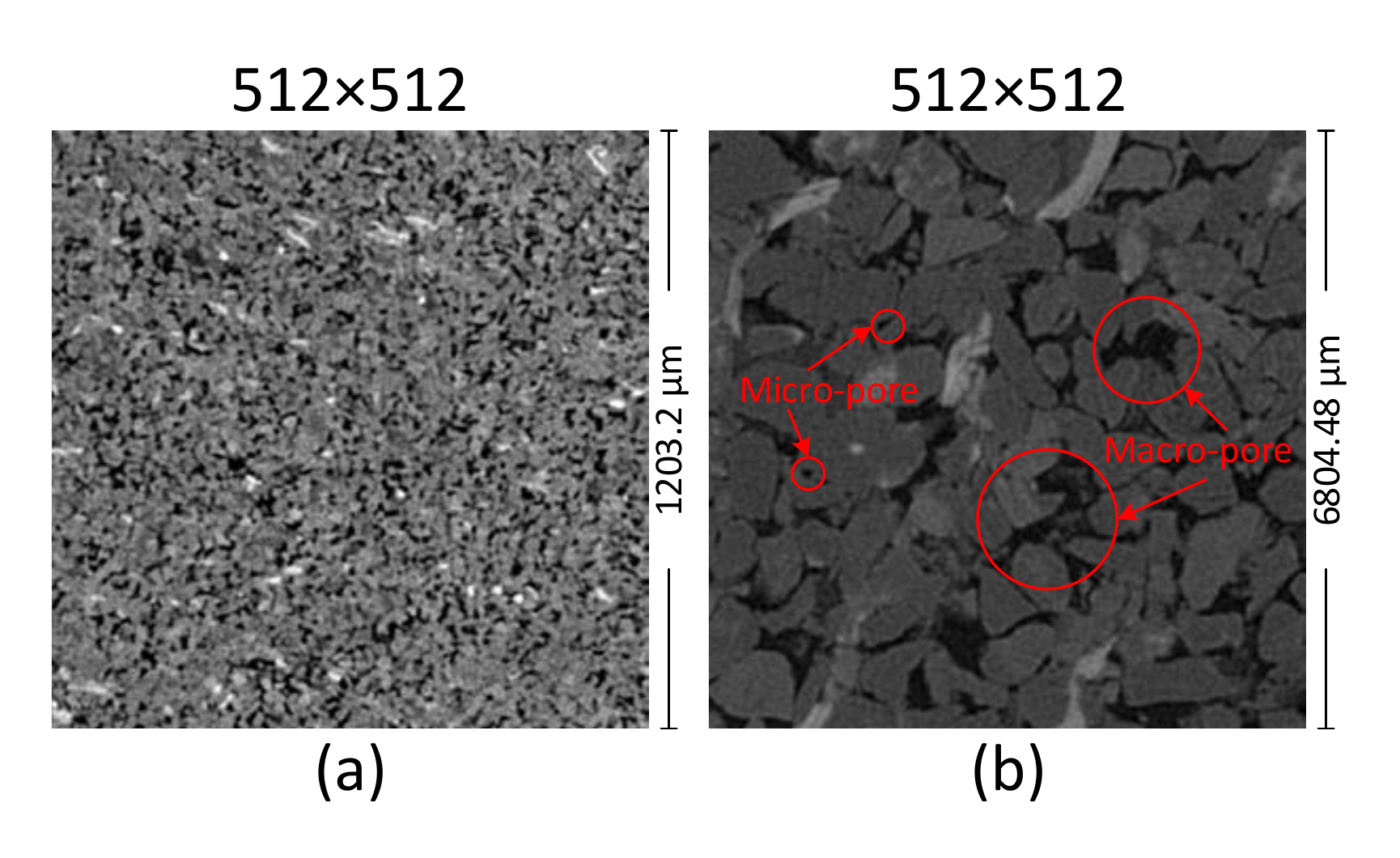}
	\caption{The micro-CT images of the core sample.}
	\label{fig:Fig1}
\end{figure*}

Numerical modeling, seepage simulation and physical properties analysis for core microstructure are important auxiliary means for oil and gas exploration and development \cite{KUANG20211,zhu2019challenges,bostanabad2018computational,ju20143d,hou20213d,ju2017multi}. Accurate modeling of core 3D pore structure is of fundamental significance to the study of physical and transport properties \cite{li2018transfer,bultreys2016imaging,tian2020surrogate,fu2022stochastic,fu2021statistical}. The digital core \cite{schluter2014image,ju20183,karimpouli2020computing,gerke2017multi,zhang2019investigation,tan2021digital,feng2020end,sanematsu2019pore} is a popular technique for modeling 3D core pore structure, which can be obtained by imaging equipment, such as X-ray Computed Tomography (micro-CT). Generally, imaging equipment captures only a single length-scale core feature in a single imaging session. Moreover, due to the tradeoff between resolution and field of view (FoV), it is difficult to obtain high-resolution core images with large-FoV that simultaneously characterize macro-pores and micro-pores \cite{li2020three}. Figure \ref{fig:Fig1} shows micro-CT images of the same core sample at different resolutions. The length-scale of (a) is 13.29$\mu \mathrm{m}$, and (b) is 2.35$\mu \mathrm{m}$, and the FoV of low-resolution image (a) is larger than that of high-resolution image (b). Macro-pores and the skeleton of pores are easily captured by low-resolution images (LRI). On the contrary, the high-resolution images (HRI) captures micro-pores and fine pore edges in addition to part of macro-pores. Compared with the high-resolution 3D pore model with large-FoV, whether 3D pore models based on LRI with large-FoV or HRI with small-FoV have limitations in describing the core structure. That is, on the one hand, HRI with small-FoV are insufficient to represent the whole core sample, and on the other hand, LRI with large-FoV lack high-scale pore features (micro-pores and fine pore edge). Although the physical properties of cores such as porosity and permeability are mainly determined by the macro-pores. Micro-pores, pore geometry and pore surface morphology are also affecting the adsorption and desorption of oil and gas \cite{biswal2009modeling,wu2020comprehensive,wu2019multiscale}. Micro-pores and fine pore edges are also important for accurate modeling of pore structure. In conclusion, imprecise pore structure modeling may introduce errors in numerical simulation. The combination of LRI with large-FoV and HRI with small-FoV to build a fine multi-scale 3D pore structure model with large-FoV is of great significance to improve the accuracy of numerical simulation and seepage analysis. 

Multiscale reconstruction \cite{fernandes1996multiscale,lin2019multiscale,ruspini2021multiscale,zhang2019efficient,jiao2008modeling} is a feasible way to model fine multi-scale 3D pore structures with large-FoV. In recent years, multiscale reconstruction has attracted tremendous attention of researchers. Yao et al \cite{yao2013construction} adopted the simulated annealing (SA) method to reconstruct the low-resolution 3D pore structure containing macro-pores, and Markov Chain Monte Carlo (MCMC) method to reconstruct the high-resolution 3D pore structure containing micro-pores, then superposed them to construct the multi-scale 3D pore structure. Marina V et al \cite{karsanina2018enhancing} proposed an image fusion method to convert CT images of different scales to the same resolution, and then superposed them to construct a multiscale soil structure model. Tabmasebi \cite{tahmasebi2018nanoscale} used similarity mapping to determine the corresponding region of a small-size fine-scale image in a large-size coarse-scale image, and then adopted HPPYS algorithm to iteratively refine the coarse image to contain fine-scale information. Wu et al \cite{wu2020comprehensive} proposed a multiscale modeling algorithm for complex rocks based on template matching, which fused different structural patterns into one image. Ji et al \cite{ji2019multiscale} adopted CCSIM-TSS algorithm to generate 3D pore structures based on different resolutions CT images, and proposed a multicomponent superposition algorithm to integrate 3D models of different components in shale. The above methods are committed to transform images of different resolutions (or scales) to the same resolution (or scale) by some technologies or algorithms, and construct multiscale model according to the designed superposition rule. However, some of them \cite{karsanina2018enhancing,okabe2007pore} are difficult to avoid errors caused by the overlap of pores or phases in images at different scales in the complicated superposition process, even with the corresponding improvement. In addition, there are methods (such as \cite{wu2020comprehensive,lin2019multiscale,tahmasebi2018nanoscale}) that mainly deal with 2D images. First, the corresponding regions of the small-scale image in the large-scale image are matched, and then the reconstruction algorithm is used to reconstruct other regions in the large-scale image. Due to the lack of spatial information, these methods only generate a 2D multiscale image. Moreover, the corresponding regions of HRI and LRI are difficult to match in practice.

Therefore, the researchers came up with an alternative way: to reconstruct 3D multiscale pore structure by introducing high-resolution (or high-scale) 2D/3D image information into 3D low-resolution (or low-scale) structures \cite{li2018accurate,wang2018semi,song2021improved,ronneberger2015u}. For instance, Li et al \cite{li2020three} proposed a method to fuse spatial information from 2D high-resolution image into a 3D low-resolution pore structure, which reconstruct multiscale pore structure with large-FoV by combining high-resolution 2D pore image with small-FoV and low-resolution 3D pore structure with large-FoV, but this method did not consider the pore geometry. Wang et al \cite{wang2018three} proposed a local-similarity statistic reconstruction (LSSR) method to reconstruct 3D high-resolution porous structure by combining a set of increasing resolution and decreasing FoV micro-CT images. With the application of deep learning in this field \cite{fu2022stochastic,mosser2017reconstruction,zhang20223d,xia2022multi,guan2021reconstructing,wang2019ct} some researchers proposed the multiscale reconstruction method based on Generative Adversarial Network (GAN). Shams et al \cite{shams2020coupled} used coupled Generative Adversarial and Auto-Encoder neural network to reconstruct 3D multiscale porous media that contains inter-grain and intra-grain pores simultaneously. Yang et al \cite{yang2022multi} used conditional Generative Adversarial Network (cGAN) to reconstruct 3D multiscale pore structure from low-resolution core images. These GAN-based techniques generate multi-scale pore structures effectively, but they also have some disadvantages such as large amount of training data and long training time. In general, although studies on multiscale reconstruction has increased, there are still some problems to be solved.

In this paper, a multiscale reconstruction algorithm based on multiple dictionaries learning is proposed. This method introduces the pattern information (macro-pore edge patterns and micro-pore patterns) of the high-resolution 3D core structure with small-FoV into the low-resolution 3D core structure with large-FoV. The introduction of high-scale patterns can modify the pore edges in low-resolution 3D core structure and adds micro-pores that cannot be captured in low-resolution images. The two structures (low- and high- resolution structure) can be extracted from different regions of the same core, which means that their locations do not have to be matched. Through experiments on simulation data (real HRI and simulation LRI data pair), the effectiveness of reconstruction results is proved from quantitative (physical properties and seepage characteristics) and qualitative (visualization of reconstruction results) comparisons. Furthermore, the conclusions of this paper are also verified by real sample experiments. 

The reminder of this paper is presented as: Section \ref{sec:Methodology} describes the methodology. The multiscale reconstruction results are presented and analyzed in Section \ref{sec:results and discussion}. In Section \ref{sec:conclusion}, we make concluding remarks.

\section{\label{sec:Methodology}Methodology}

\begin{figure*}[h]
	\centering
	\includegraphics[keepaspectratio=true, width=0.45\textwidth]{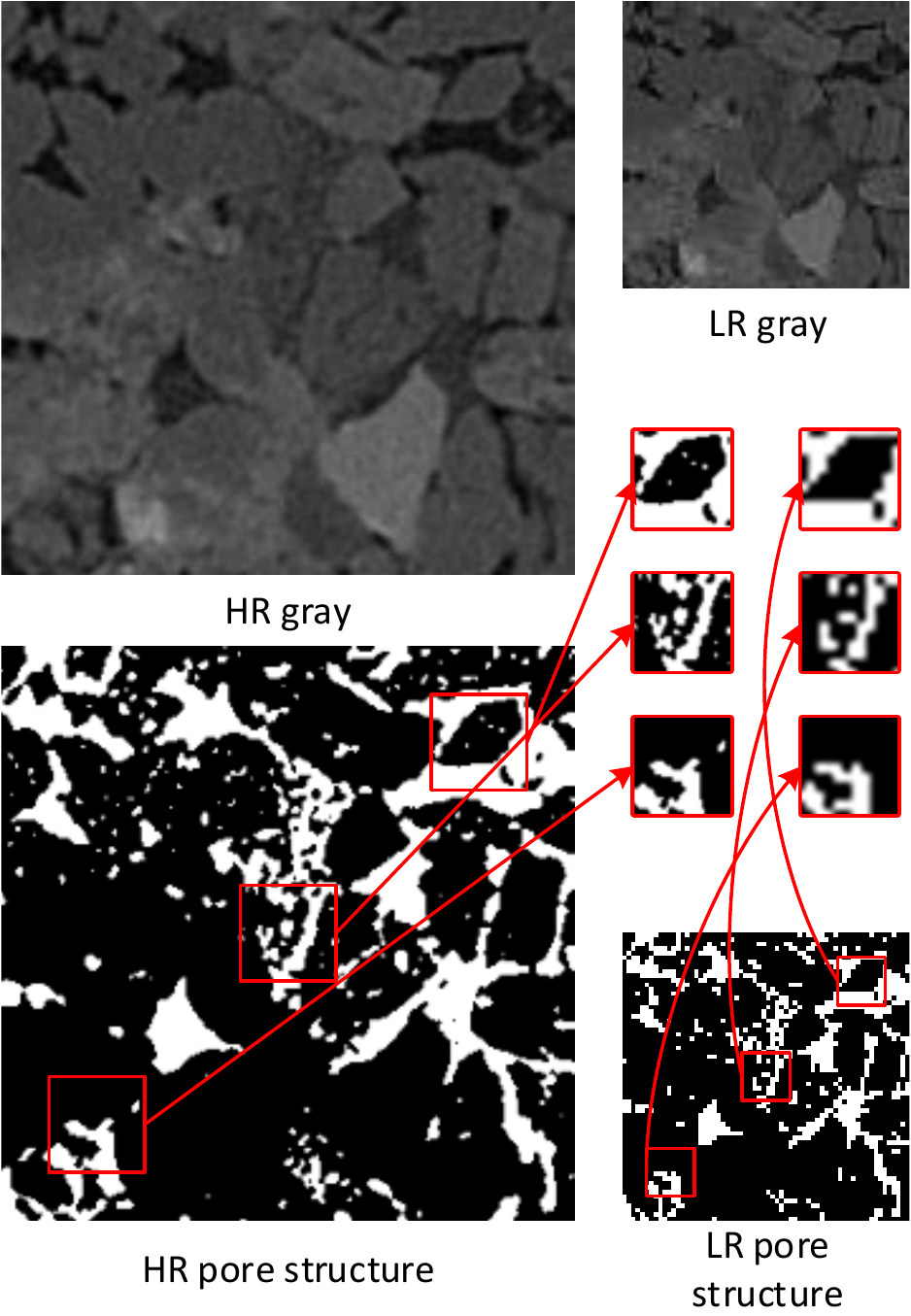}
	\caption{Different resolutions 2D core images and corresponding pore structure.} 
	\label{fig:Fig2}
\end{figure*}

To realize the multiscale reconstruction, both FoV and length-scale of image should be considered. The relationship of core image FoV, length-scale and image size are as follows: 
\begin{small}
	\begin{eqnarray} 
	\textit{iFOV}=\textit{iL} \times \textit{iS}
	\end{eqnarray} 
\end{small}
where \textit{iFoV} is FoV of image, \textit{iL} is length-scale and \textit{iS} is image size. Length-scale \textit{iL} decreases as the image resolution increases. Different resolutions 2D core images and corresponding pore structure are shown Figure \ref{fig:Fig2}. The low-resolution gray image is obtained by downsampling the high-resolution gray image. In the low-resolution pore structure, the main structure of most pores is preserved. Red boxes mark the details of the two images, which reflects the difference in the local patterns of the two images. High-resolution pattern contains edge and micro-pore, and low-resolution pattern contains pore skeleton. Although some image processing algorithms (such as image upsampling) can improve image resolution (reduce the \textit{iL}), it is difficult to accurately model fine multiscale pore structures without imaging with high-resolution equipment. This is because low-resolution gray images do not capture details features such as micro-pores and fine pore edges. It is instructive that multiscale reconstructions can be seen as introducing details features from HRI with small-FoV into the enhanced resolution LRI with large-FoV for joint modeling.

\begin{figure*}[htbp]
	\centering
	\includegraphics[keepaspectratio=true, width=0.7\textwidth]{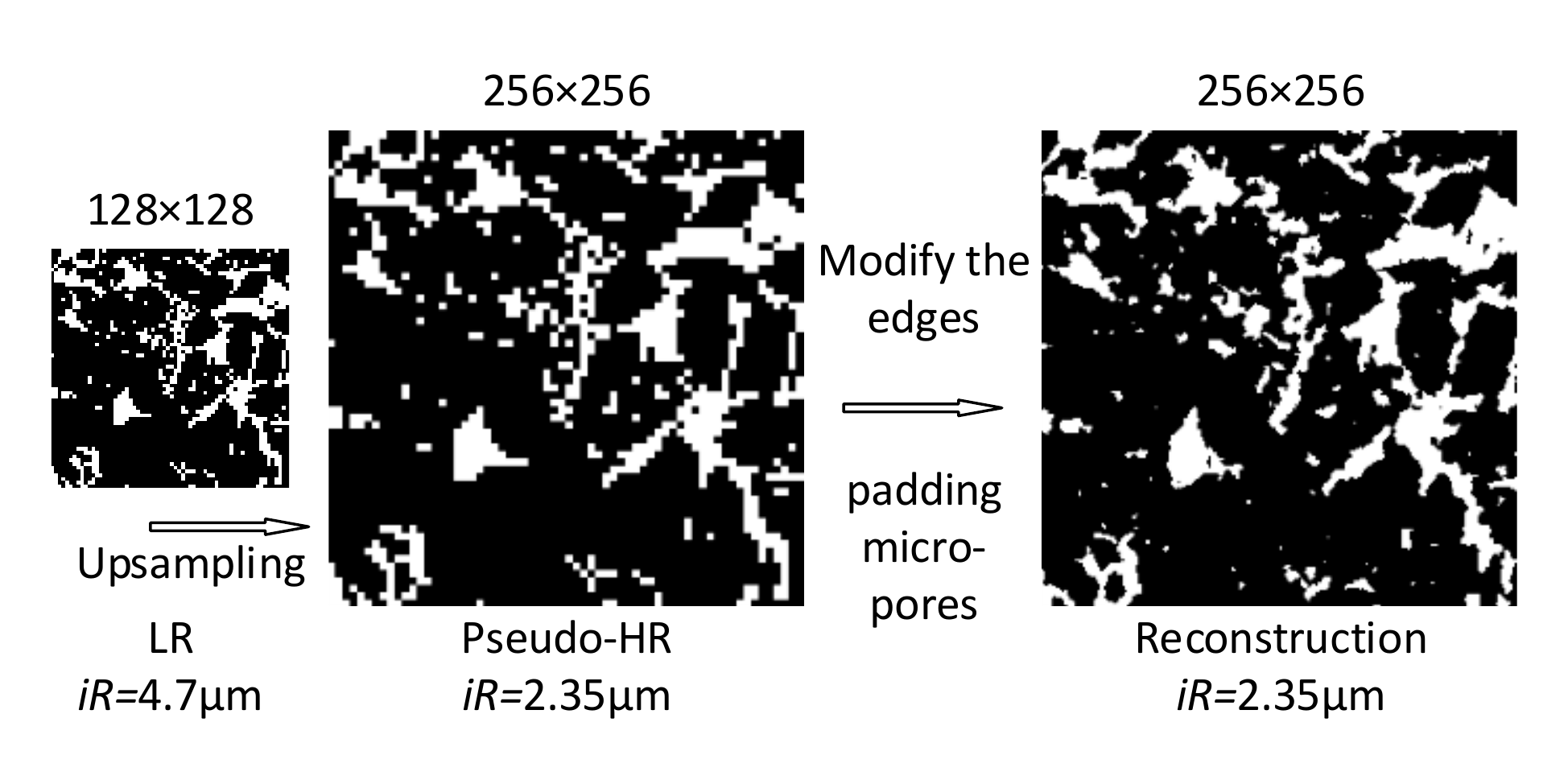}
	\caption{The flowchart of multiscale reconstruction.} 
	\label{fig:Fig3}
\end{figure*}

At the micro scale, cores have local similarity. Arguably, if the main structure of two patterns is similar on the same scale, then their edges are also similar \cite{wang2018three,li2021reconstruction,ding2018improved}. Even from different regions of the same core, LRI pore edges can be effectively modified as long as HRI contains sufficient information. Therefore, based on the above assumptions, multi-scale reconstruction can be simply divided into the following steps (as shown in Figure \ref{fig:Fig3}):

\begin{itemize}
	\item [1)]
	Enhance LR pore structure resolution by upsampling (reduce \textit{iL} of LRI);;
	\item [2)]
	Modify edges by matching LR and HR patterns of the same scale;
	\item [3)]
	Padding micro-pores that LR cannot capture.
\end{itemize}

This paper builds edge pattern dictionary (\textit{EPD}) and micro-pore dictionary (\textit{MPD}) from HRI. It is worth noting that multiple dictionaries of edge patterns with different scales are built. The pore edges of LRI are modified in stages by matching the patterns in different scale edge pattern dictionaries. Then, the micro-pore patterns from the micro-pore dictionary are random padded into LRI. 

\subsection{Upsampling Mapping}

\begin{figure*}[h]
	\centering
	\includegraphics[keepaspectratio=true, width=0.5\textwidth]{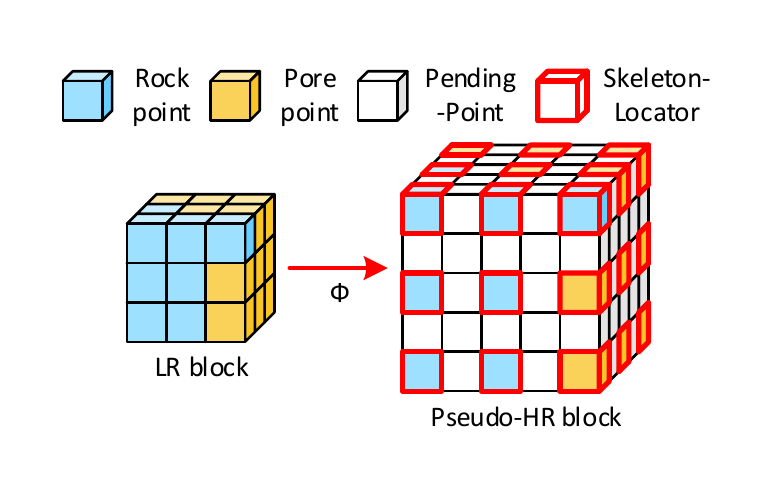}
	\caption{The Upsampling Mapping.} 
	\label{fig:Fig4}
\end{figure*}

Traditional image upsampling can increase the \textit{iS} and reduce the \textit{iL} by interpolating the original pixels. As shown in Figure \ref{fig:Fig4}, this paper defines an Upsampling Mapping $\Phi$:
\begin{small}
	\begin{eqnarray} 
	\text {pseudo-HR} =\Phi(\textit {LRI})
	\end{eqnarray} 
\end{small}
The pseudo-HR block consists of Skeleton-Locator (\textit{SL}) and Pending-Points. Pixels of LR block (include rock points and pore points) are taken as the Skeleton-Locator to guidance the neighboring pixels (Pending-Points). The Skeleton-Locator is the framework for the pattern, which represents the main structure of the pores. Pending-Points reflect the development of pore edges, which are determined by matching similar Skeleton-Locator in edge pattern dictionary. 

Let the length-scale of LRI be $\textit{iL}_{\textit{LR}}$, and after $n$ times Upsampling Mapping is performed:
\begin{small}
	\begin{eqnarray} 
	iL_{\textit{LR}}^{n}=\frac{\textit{iL}_{\textit{LR}}}{2^{n}}, n=0,1,2, \ldots
	\end{eqnarray} 
\end{small}
The maximum value of $n$ should meet the following conditions:
\begin{small}
	\begin{eqnarray} 
	\frac{\textit{iL}_{\textit{LR}}}{2^{n+1}} \leq \textit{iL}_{\textit{HR}} \leq \frac{\textit{iL}_{\textit{LR}}}{2^{n}}
	\end{eqnarray} 
\end{small}
Theoretically, the resolution of multiscale structures with the highest resolution that can be reconstructed by the algorithm can be determined according to Inequation (4). The multiscale structure highest resolution is $\frac{\textit{iL}_{\textit{LR}}}{2^{n}}$ when $n$ is maximum.

\subsection{Multiple Scale of Edge Pattern Dictionaries}

\begin{figure*}[h]
	\centering
	\includegraphics[keepaspectratio=true, width=0.75\textwidth]{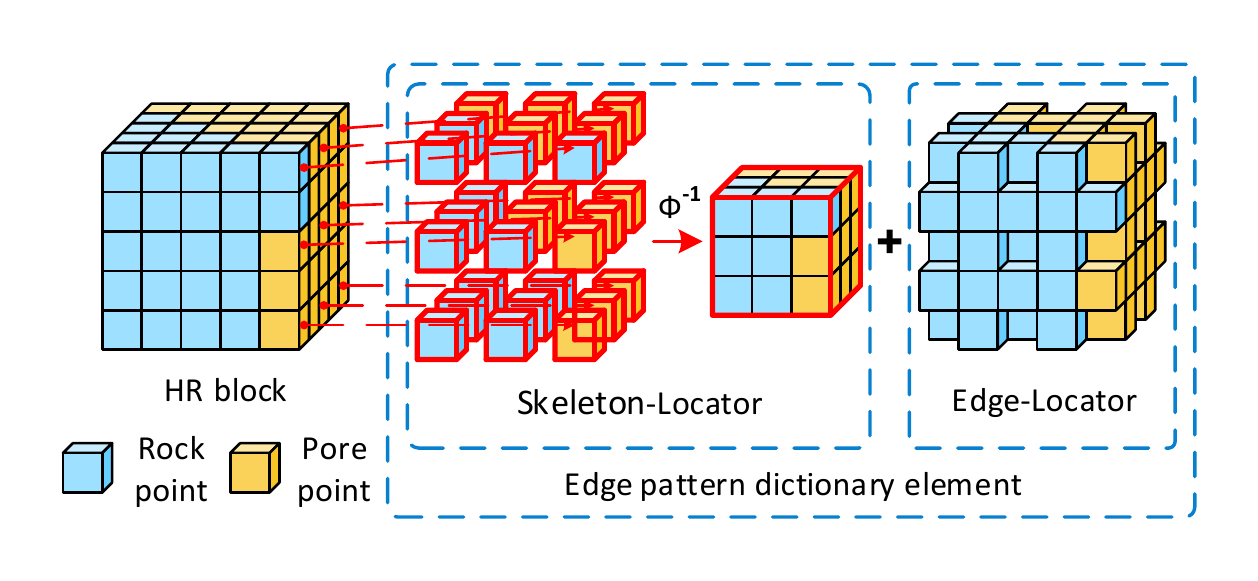}
	\caption{The edge pattern dictionary element.} 
	\label{fig:Fig5}
\end{figure*}

Figure \ref{fig:Fig5} shows the edge pattern dictionary element. In this paper, a $5^3$-size template is used to scan HRI to build edge pattern dictionary. The edge pattern dictionary element consists Skeleton-Locator ($\textit{EPD}_\textit{SL}$) and Edge-Locator ($\textit{EPD}_\textit{EL}$). $\Phi^{-1}$ is used to extract Skeleton-Locator from HR block:
\begin{small}
	\begin{eqnarray} 
	\textit{EPD}_\textit{SL}=\Phi^{-1}(\text {HR block})
	\end{eqnarray} 
\end{small}
where ($\textit{EPD}_\textit{SL}$) and ($\textit{EPD}_\textit{EL}$) capture the main structure and rich edge details of the patterns respectively. Removing redundant patterns, a single scale edge pattern dictionary is expressed as:
\begin{small}
	\begin{eqnarray}
	\textit{EPD}=\sum_{i} \textit{EPD}_{\textit{SL}_{i}}\left(\sum_{ij} \textit{EPD}_{\textit{EL}_{ij}}\right),(i, j=1,2, \ldots)
	\end{eqnarray}
\end{small}
The $\textit{EPD}_{\textit{SL}_{i}}$ indicates the $i$-th class Skeleton-Locator in the \textit{EPD}. The $\textit{EPD}_{\textit{EL}_{ij}}$ indicates the $j$-th Edge-Locator in $i$-th class Skeleton-Locator. Patterns from pores with different scales may have the same Skeleton-Locator, so a single scale \textit{EPD} will confuse their Edge-Locator and leads to a mismatch. It is corrected by building multiple edge pattern dictionaries of different scales. 

In general, image downsampling decreases the resolution while loses the high-scale features and retains the low-scale features. Therefore, a lower-scale pattern dictionary can be built by scanning the downsampled HRI. Let the length-scale of HRI be $\textit{iL}_{\textit{HR}}$. The edge dictionary built by HRI is $\textit{EPD}^1$. The HRI is downsamped $m$ times, and the resolution of HRI becomes 
\begin{small}
	\begin{eqnarray} 
	iL_{\textit{HR}}^{m}=2^{m} \cdot \textit{iL}_{\textit{HR}}, \quad m=0,1,2, \ldots
	\end{eqnarray}
\end{small}
The corresponding edge pattern dictionary is $\textit{EPD}^m$. The maximum value of $m$ should meet the following conditions: 
\begin{small}
	\begin{eqnarray} 
	\left\{\begin{array}{c}
	2^{m} \cdot \textit{iL}_{\textit{HR}} \leq \textit{iL}_{\textit{LR}} \leq 2^{m+1} \cdot iL_{\textit{HR}} \\
	\frac{\textit{iS}_{\textit{HR}}}{2^{m}} \geq \text { template size. }
	\end{array}\right.
	\end{eqnarray}
\end{small}
When $m$ is the maximum value, it means that the feature scale of HRI is similar to LRI after $m$ times of downsampling. 

\subsection{Edge Pattern Matching and multi-stage reconstruction}

\begin{figure*}[h]
	\centering
	\includegraphics[keepaspectratio=true, width=1\textwidth]{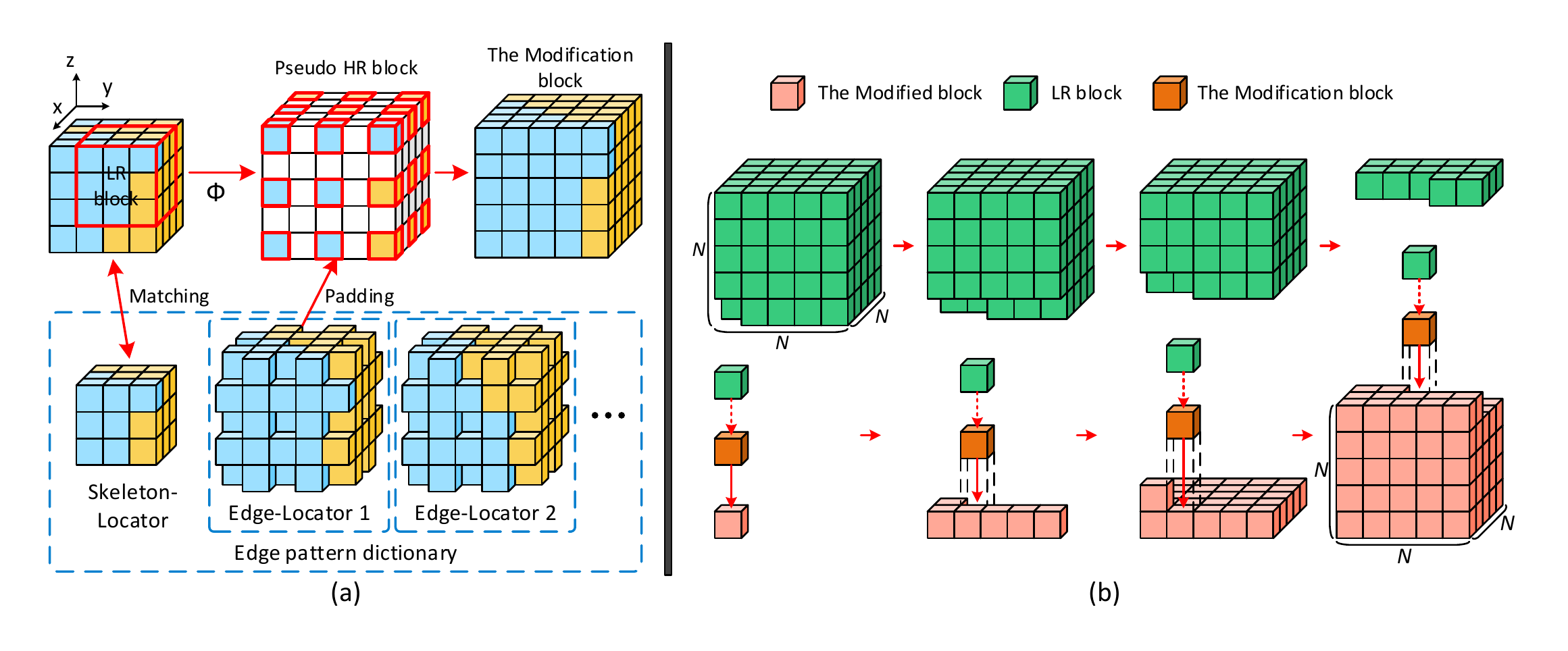}
	\caption{(a) shows the process of edge pattern matching. (b) is the edge modification process.} 
	\label{fig:Fig6}
\end{figure*}

Figure \ref{fig:Fig6} vividly illustrates the single-stage reconstruction process. Considering the connectivity of pores, the matching mechanism is divided into two steps. First, $\textit{LR}_{\textit{Block}}$ and $\textit{EPD}_\textit{SL}$ are matched according to the following formula:
\begin{small}
	\begin{eqnarray} 
	\arg \min \left(\left|\textit{LR}_{\textit{Block}}(x, y, z)-\textit{EPD}_{\textit{SL}_{i}}(x, y, z)\right|\right), i=1,2, \ldots
	\end{eqnarray}
\end{small}
Formula (9) is used to match the most similar $i$-th class Skeleton-Locator in the \textit{EPD}. Then, the Edge-Locator with the best connectivity is matched according to the following formula:
\begin{small}
	\begin{eqnarray} 
	\arg \min \left(\sum_{r}\left|\operatorname {\emph {Mod}}_{ij}^{(r)}(x, y)-\operatorname {\emph {Adj}}^{(r)}(x, y) \right|\right), j=1,2, \ldots
	\end{eqnarray}
\end{small}
The $\emph {Mod}_j$ is the Modification block padded with $j$-th Edge-Locator that in $i$-th class Skeleton-Locator. The $\emph {Mod}^{(r)}$ indicates the $r$-th direction surface of \emph {Mod} and the $\emph {Adj}^{(r)}$ is the corresponding adjacent surface is in the modified block. 

In order to reconstruct the edge that is more similar to the real edge, the reconstruction of the edge is multi-stage. The LRI pore edges are reconstructed in stages by matching edge pattern dictionaries that is low-scale to high-scale. Let $m$ be the maximum. The $\emph {Rec}(\mathrm{,})$ indicates the single-stage reconstruction process. So, the multi-stage reconstruction can express as:
\begin{small}
	\begin{eqnarray} 
	\left\{\begin{array}{l}
	\emph {PMS}^{1}=\operatorname{\emph {Rec}}\left(\emph {LRI}, \emph {EPD}^{m}\right) \\
	\emph {PMS}^{2}=\operatorname{\emph {Rec}}\left(\emph {PMS}^{1}, \emph {EPD}^{m-1}\right) \\
	\ldots \\
	\emph {PMS}^{m}=\operatorname{\emph {Rec}}\left(\emph {PMS}^{m-1}, \emph {EPD}^{1}\right)
	\end{array}\right.
	\end{eqnarray}
\end{small}
The $\emph {PMS}^m$ represents the multiscale structure that reconstructed by $m$-stage \emph {EPD}s modification edge. After multi-stage edge reconstruction, the $iFoV$ and $iL$ of the $\emph {PMS}^m$ are as follows:
\begin{small}
	\begin{eqnarray} 
	\begin{aligned}
	\textit{iFoV}_{\emph {PMS}^{m}} &=\textit{iFoV}_{\emph {LR}} \\
	iL_{\emph {PMS}^{m}} &=\frac{\textit{iL}_{\emph {LR}}}{2^{m}}
	\end{aligned}
	\end{eqnarray}
\end{small}

\subsection{Micro-pore pattern dictionary and micro-pore padding}

Besides the edge reconstruction, the padding of micro-pores is also important. The smallest Connected Component (\emph {CC}) that can be counted in $\emph {PMS}^m$ is a single pixel. The corresponding number of pixels in $\emph {PMS}^m$ is $\left(\left\lceil\frac{iL_{\emph {PMS}}^m}{iL_{\emph {HR}}}\right\rceil\right)^{3}$. Moreover, the corresponding number of pixels in HRI is $\left(\left\lceil\frac{iL_{\emph {LR}}}{iL_{\emph {HR}}}\right\rceil\right)^{3}$. Therefore, the Connected Component of HRI ($\emph {CC}_{iL_{\emph {HR}}}$) contains micro-pores that not be captured by $\emph {PMS}^m$. The range of $\emph {CC}_{iL_{\emph {HR}}}$ is following:
\begin{small}
	\begin{eqnarray} 
	\mathrm{\emph {CC}}_{iL_{\emph {HR}}} \subseteq\left[\left(\left\lceil\frac{iL_{\emph {PMS}_m}}{iL_{\emph {HR}}}\right\rceil\right)^{3},\left(\left\lceil\frac{iL_{\emph {LR}}}{iL_{\emph {HR}}}\right\rceil\right)^{3}\right]
	\end{eqnarray}
\end{small}
Two-Pass method \cite{shapiro1996connected} is a classical image processing algorithm to mark the Connected Component (\emph {CC}) of binary image. In this paper, the Two-Pass method is adopted to mark \emph {CC}s in the range of $\left[\left(\left\lceil\frac{iL_{\emph {PMS}_m}}{iL_{\emph {HR}}}\right\rceil\right)^{3},\left(\left\lceil\frac{iL_{\emph {LR}}}{iL_{\emph {HR}}}\right\rceil\right)^{3}\right]$ in HRI to build a micro-pore pattern dictionary. \emph {MPD} elements are extracted by 3D Bounding Rectangler of these Connected Components. 

\begin{figure*}[h]
	\centering
	\includegraphics[keepaspectratio=true, width=0.50\textwidth]{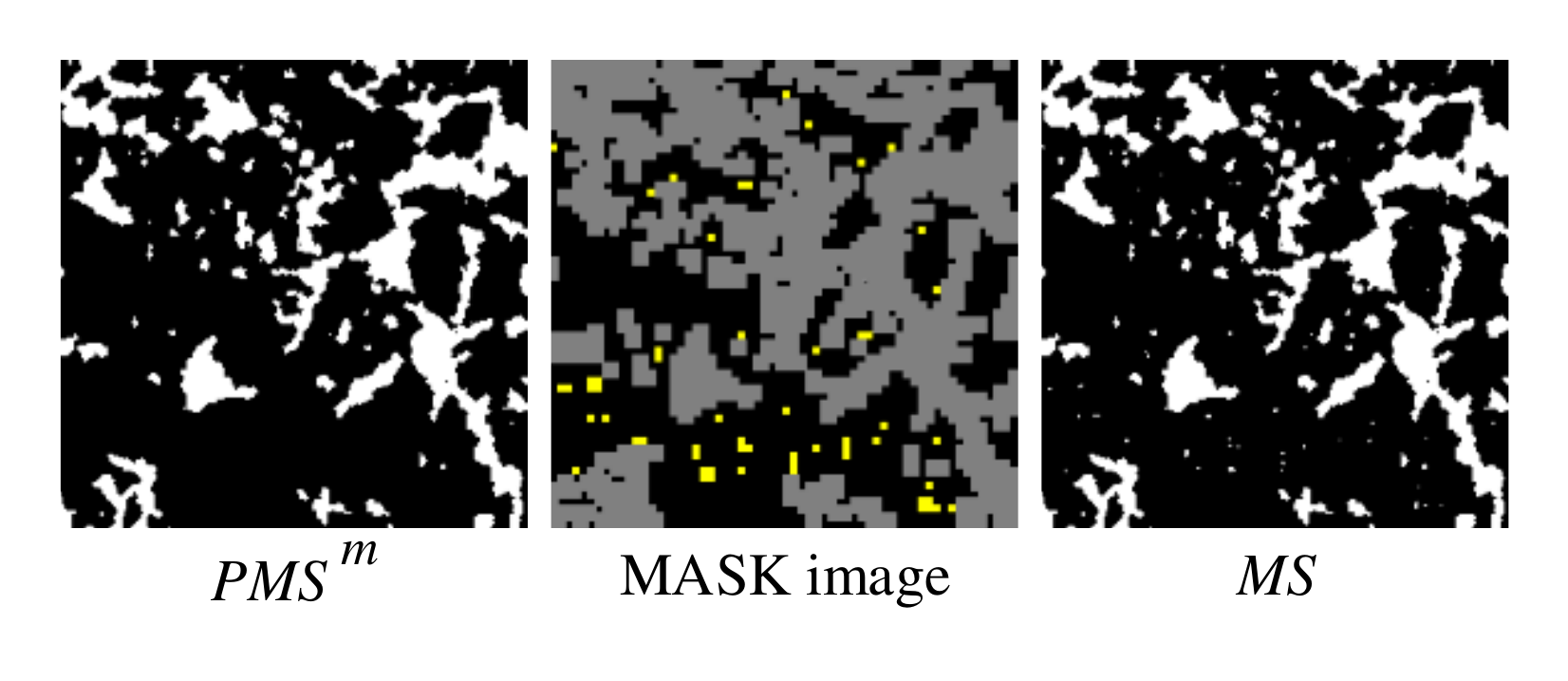}
	\caption{Micro-pore padding process.} 
	\label{fig:Fig7}
\end{figure*}

To avoid overlap, Connected Component in the $\emph {PMS}^m$ is marked with MASK. The MASKs are calculated by morphological dilation operation. To facilitate understanding, the 2D image in Figure \ref{fig:Fig7} as an example. Connected Components in the $\emph {PMS}^m$ is marked (the gray area of Mask image), and then \emph {MPD} elements (the yellow area of Mask image) are randomly padded into the non-pore area (the black area of $\emph {PMS}^m$) to form a fine multiscale pore structure (\emph {MS}).

In particular, micro-pores in image are composed of two types of pores: a) Intragranular pores; b) Fragmentary intergranular pores caused by the lower resolution imaging. These intergranular pores are formed by the dense arrangement of rock particles. At the microscopic scale, they are relatively evenly distributed throughout the rock sample. Therefore, the number distribution of micro-pores in the reconstructed \emph {MS} is similar to that of HRI. The FoV of \emph {MS} is $\left(\frac{iFoV_{\emph {LR}}}{iFoV_{\emph {HR}}}\right)^{3}$ times that of HR pore structure. Thus, based on the similarity homology and statistical similarity, the elements of \emph {MPD} are repeatedly padded $\left(\frac{iFoV_{\emph {LR}}}{iFoV_{\emph {HR}}}\right)^{3}$ times in $\emph {PMS}^m$ to approximate the micro-pore distribution of the HR pore structure.

\section{\label{sec:results and discussion}Result and discussion}

\begin{figure*}[h]
	\centering
	\includegraphics[keepaspectratio=true, width=0.6\textwidth]{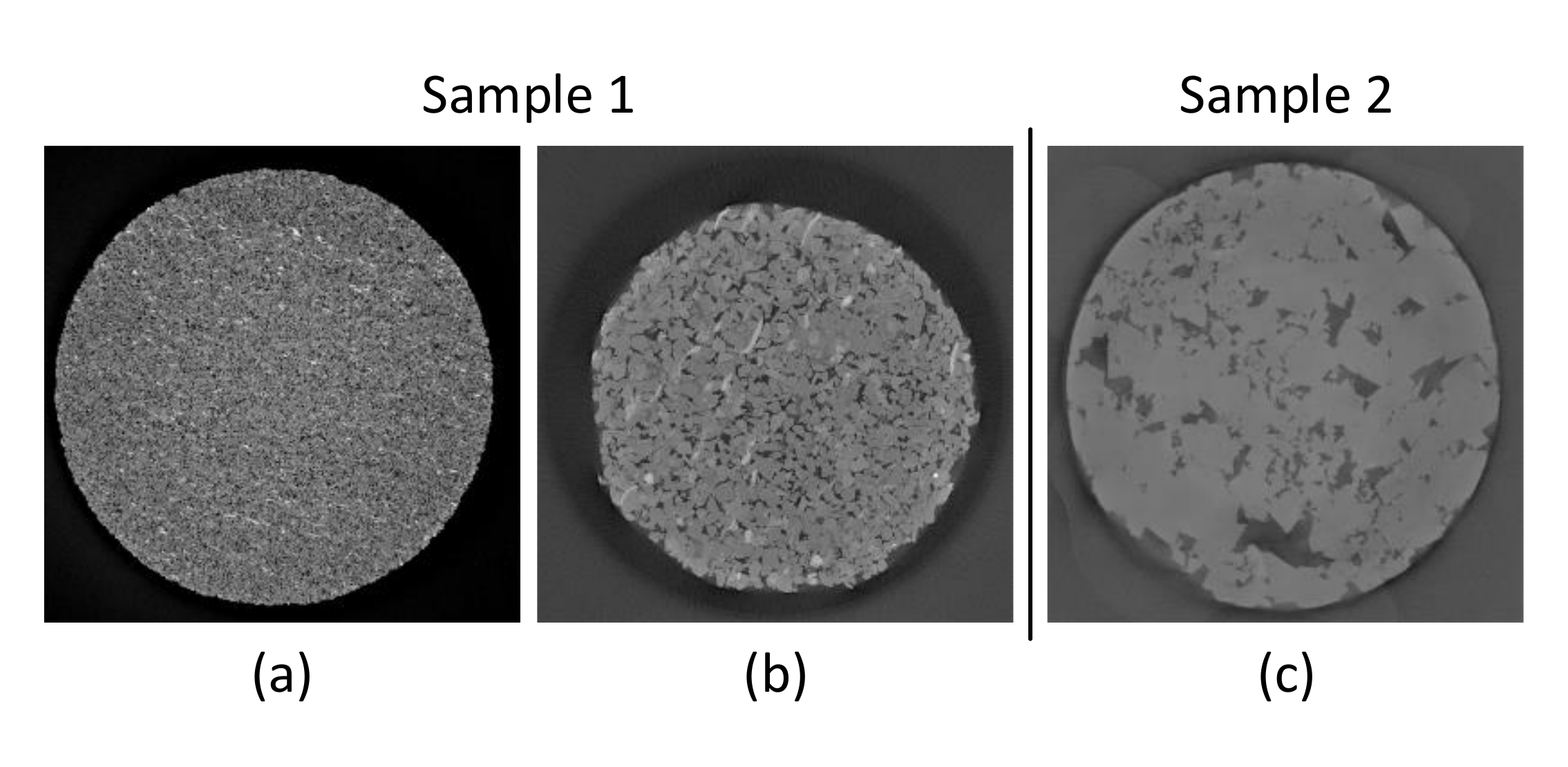}
	\caption{Example of CT images of experimental core sample} 
	\label{fig:Fig8}
\end{figure*}

In this study, simulation data experiment and real data experiment are designed to prove the effectiveness of the proposal algorithm by comparing quantitative (physical properties, seepage characteristics) and qualitative (visualization of reconstruction results). As shown in Figure \ref{fig:Fig8}, experimental core samples include sample 1 and sample 2. The length-scale of (a) is 2.35$\mu \mathrm{m}$ and (b) is 13.29$\mu \mathrm{m}$. (a) and (b) are obtained by imaging 2.5$\mathrm{cm}$ and 3$\mathrm{mm}$ (diameter) cores in sample 1, respectively. The (c)'s length-scale is 1$\mu \mathrm{m}$, which is obtained by imaging 2$\mathrm{mm}$ (diameter) core in sample 2.

\subsection{\label{section:3.1}Simulation data experiment}

The simulation data experiments are designed to facilitate comparison with real 3D high-resolution pore structures. The simulation data is processed as follows: First, two volumes of the same size are cut out of the CT image. Then, one of the them is used as the HRI, and the other is down-sampled by bicubic interpolation \cite{carlson1985monotone} as the simulation LRI. Finally, their pore structure is extracted using Otsu method \cite{trier1995evaluation}. Simulation data experiments are proceeded on sample 1(b) and sample 2(c). 

\subsubsection{Simulation data experiment for Sample 1(b)}

\begin{figure*}[h]
	\centering
	\includegraphics[keepaspectratio=true, width=0.45\textwidth]{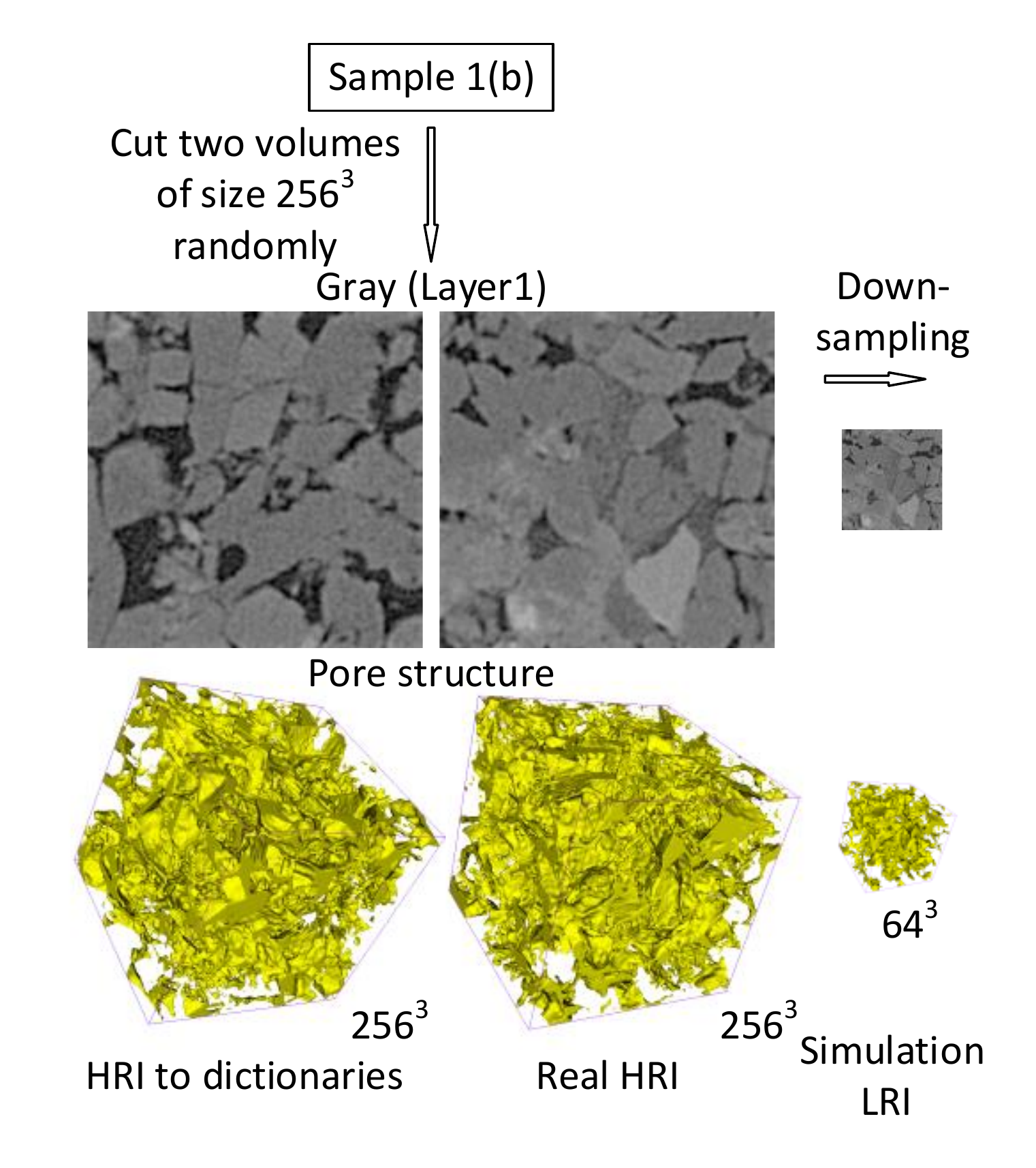}
	\caption{Simulation data for sample1(b).} 
	\label{fig:Fig9}
\end{figure*}

\begin{figure*}[htbp]
	\centering
	\includegraphics[keepaspectratio=true, width=0.9\textwidth]{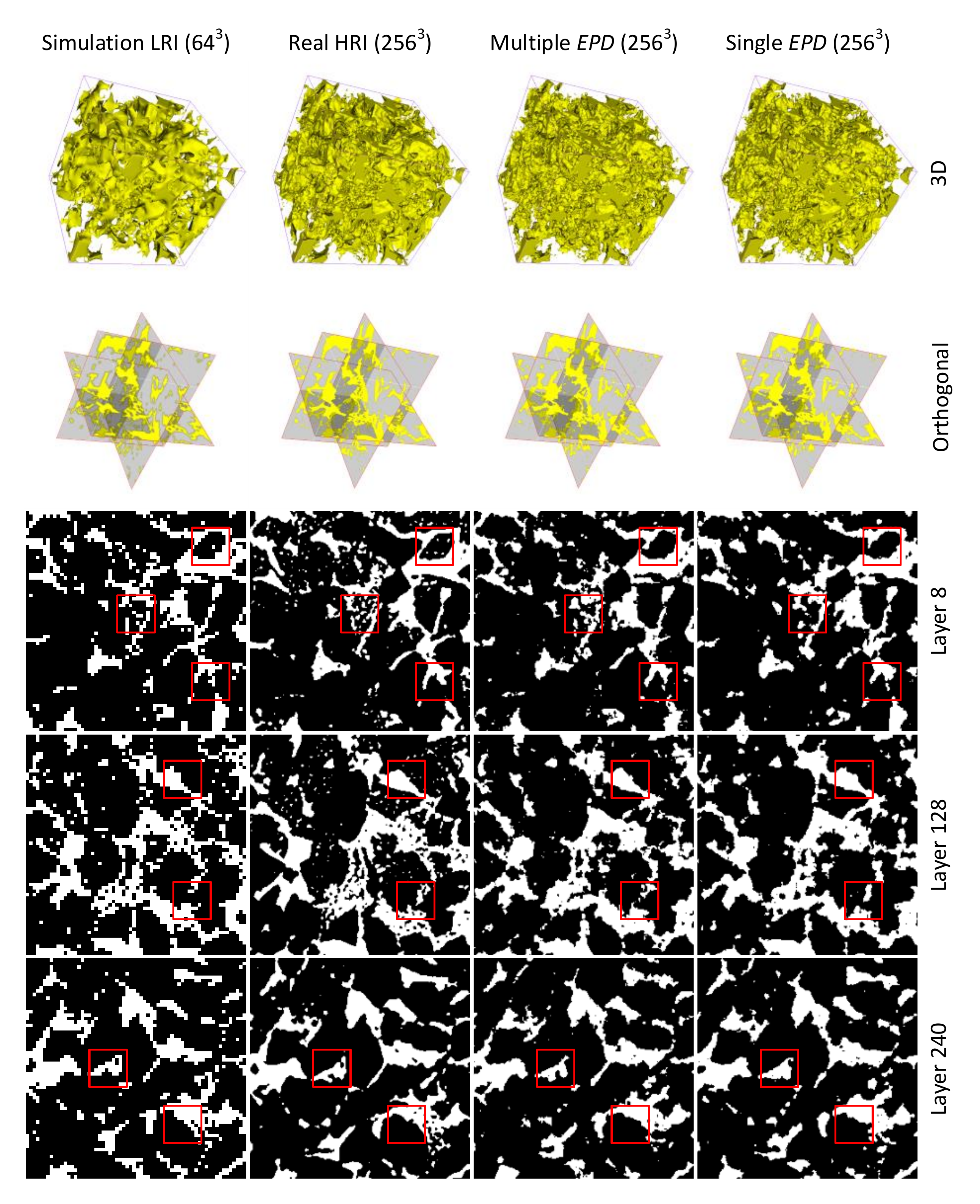}
	\caption{The qualitative comparison of the reconstructions with the real HRI on sample 1(b) simulation data experiment.} 
	\label{fig:Fig10}
\end{figure*}

Simulation data for sample 1(b) are produced following the processing in section \ref{section:3.1}, as shown in Figure \ref{fig:Fig9}. Thus, the length-scale of the HRI is 2.35$\mu \mathrm{m}$ and the simulation LRI is 9.4$\mu \mathrm{m}$. The size of HRI is $256^3$ and the simulation LRI is $64^3$. Multiple dictionaries (include multiple edge dictionaries and micro-pore dictionary) are built by HRI. According formula (8), the simulation LRI should be refined edges by a 2-stage edge reconstruction. Figure \ref{fig:Fig10} shows the qualitative comparison of reconstructions with the original HRI. To prove the effectiveness of the multiscale edge dictionaries, the results of reconstruction using single edge dictionary are compared. To prove the stability of the algorithm, multiple \emph {EPD} and single \emph {EPD} are using to reconstructed three times respectively. Figure \ref{fig:Fig10} only shows the results of one reconstruction.

According the 3D and Orthogonal row in Figure \ref{fig:Fig10}, the simulation LRI, multiple \emph {EPD}-1 and single \emph {EPD}-1 skeletons are similar to the real HRI, indicating that the multi-stage edge reconstruction preserves pore skeleton features. Furthermore, the 3D structures of both multiple \emph {EPD}-1 and single \emph {EPD}-1 are padded with a number of micro-pores that cannot be captured by simulation LRI. In Layer rows of Figure \ref{fig:Fig10}, the pore edges of the simulation LRI are rough, by multi-stage edge reconstruction, the multiple \emph {EPD}-1 and single \emph {EPD}-1 edges are both refined. However, compared to single \emph {EPD}-1, the pore edges of multiple \emph {EPD}-1 are more similar to real HRI, because multiple \emph {EPD}-1 matches the corresponding scale edge patterns in \emph {EPD}s at each stage of edge reconstruction. 

\begin{figure*}[h]
	\centering
	\includegraphics[keepaspectratio=true, width=0.41\textwidth]{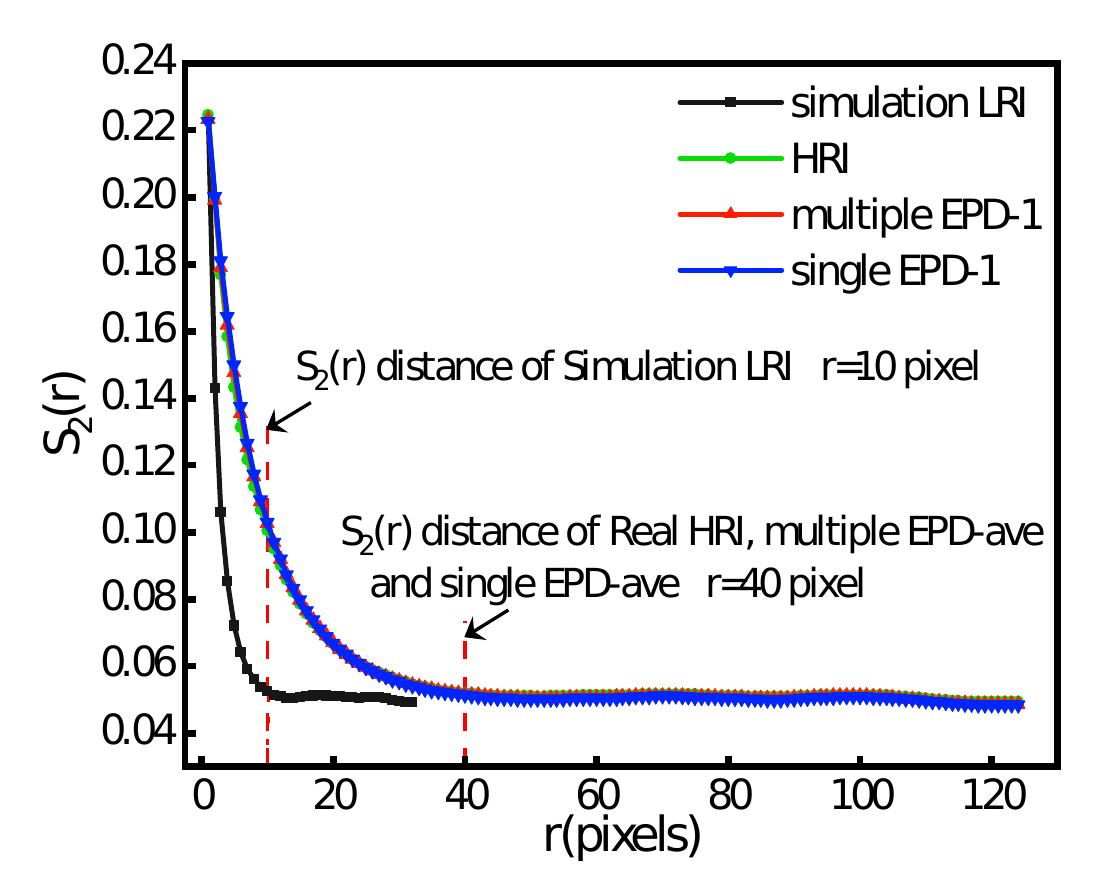}
	\caption{The comparison of the two-point correlation function ($S_2$) on sample 1(b) simulation data experiment.} 
	\label{fig:Fig11}
\end{figure*}

In Figure \ref{fig:Fig11}, the two-point correlation function ($S_2$) is used to evaluate the spatial correlation of the simulation LRI, reconstructions and real HRI. multiple \emph {EPD}-ave, single \emph {EPD}-ave represent the average of three reconstructions using multiple \emph {EPD} and single \emph {EPD} respectively. The two-points correlation distance of reconstructions ($r\approx$ 40 pixel) is equal to that of the real HRI, which is 4 times that of the simulation LRI ($r\approx$ 10 pixel). The reconstructions are consistent with the $S_2$ of the real structure, due to the multiscale reconstruction does not alter the skeleton of the pore structure. The $S_2$ of reconstructions is almost identical to the real HRI, demonstrating that the proposal method is stable. However, since the spatial correlation is mainly determined by the large pores (skeleton), $S_2$ cannot reflect changes in the pore edges. 

\begin{figure}[htbp]
	\centering
	\subfigure[pore radius distribution]{
		\includegraphics[keepaspectratio=True, width=0.42\textwidth]{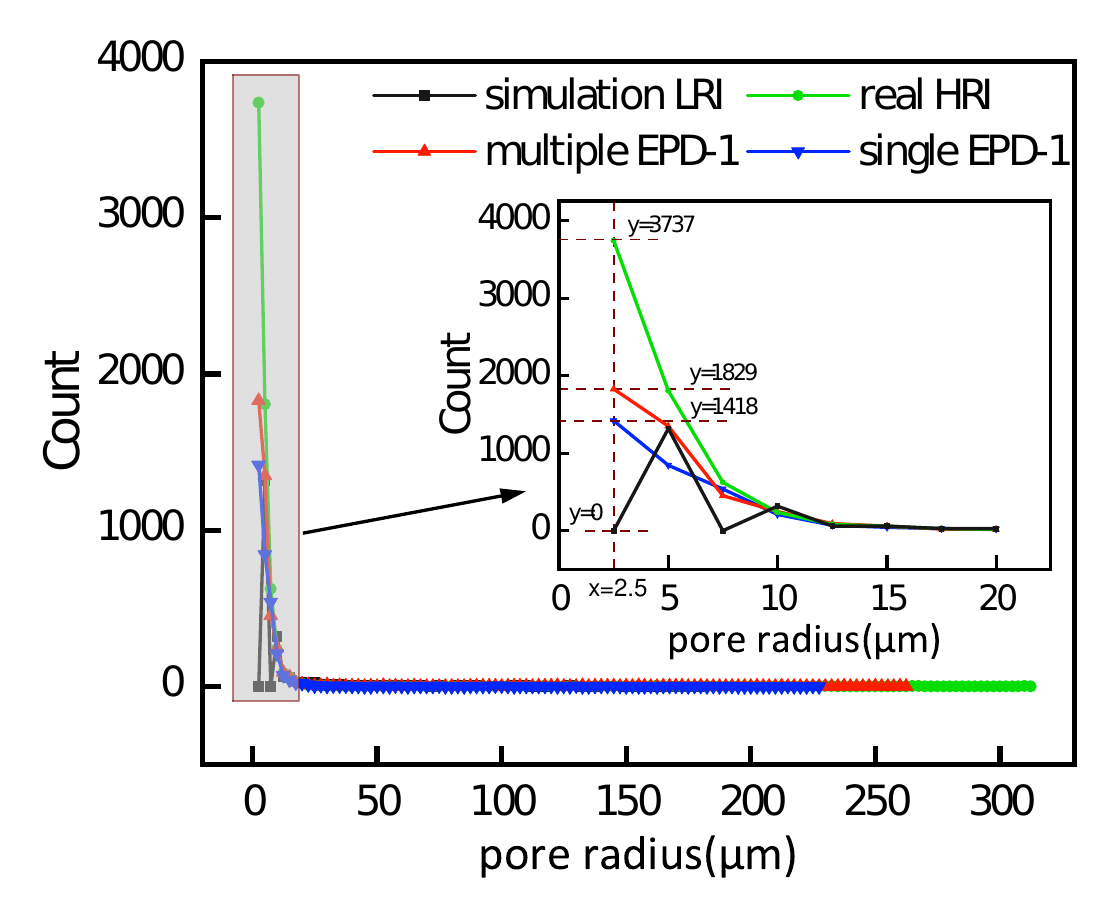}}
	\subfigure[throat radius distribution]{
		\includegraphics[keepaspectratio=True, width=0.42\textwidth]{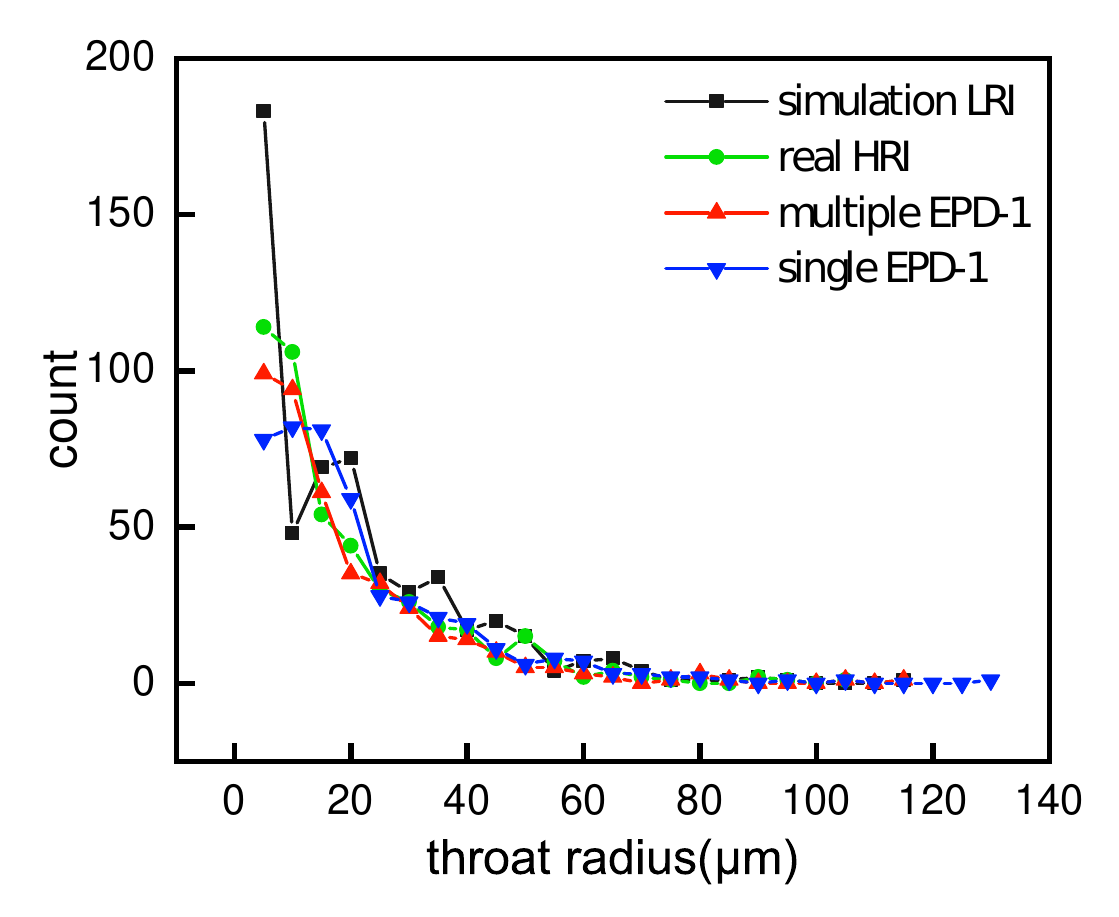}}
	\subfigure[shape factor distribution]{
		\includegraphics[keepaspectratio=True, width=0.42\textwidth]{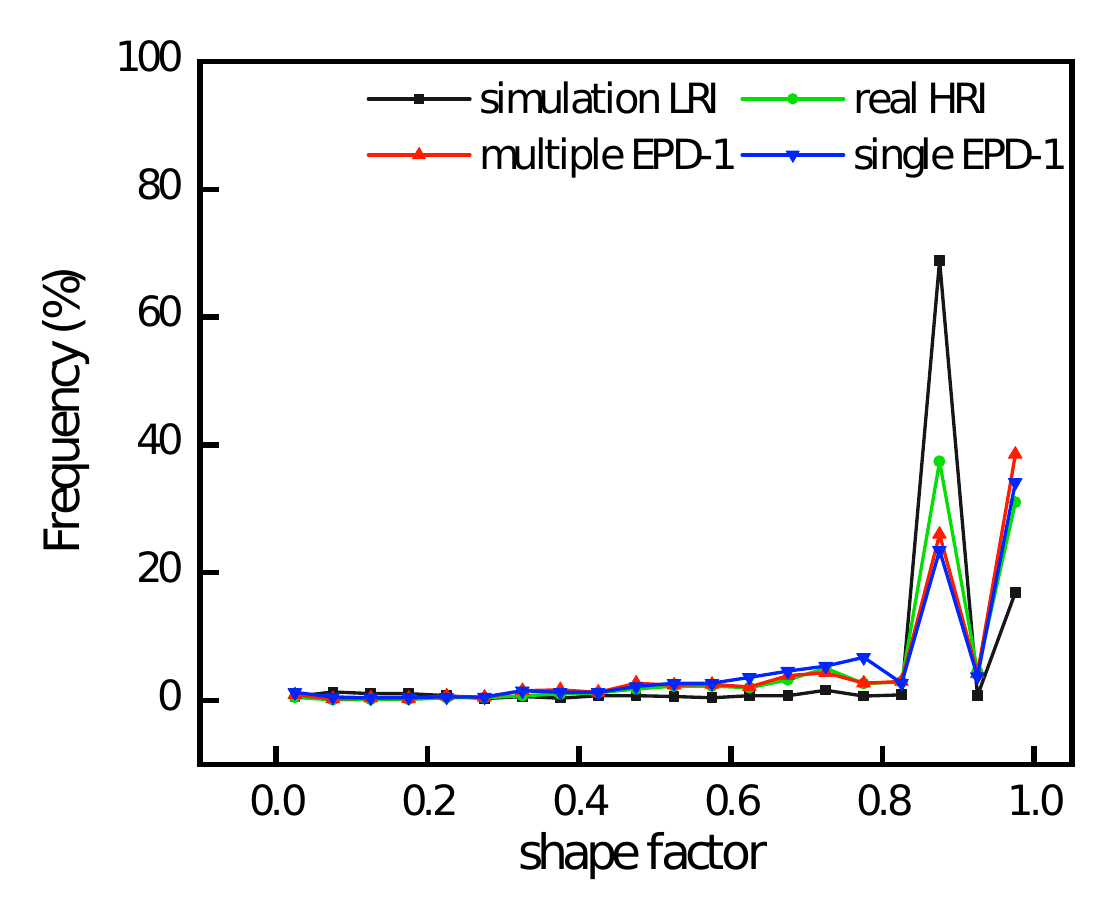}}
	\subfigure[pore shape factor-pore radius distribution]{
		\includegraphics[keepaspectratio=True, width=0.41\textwidth]{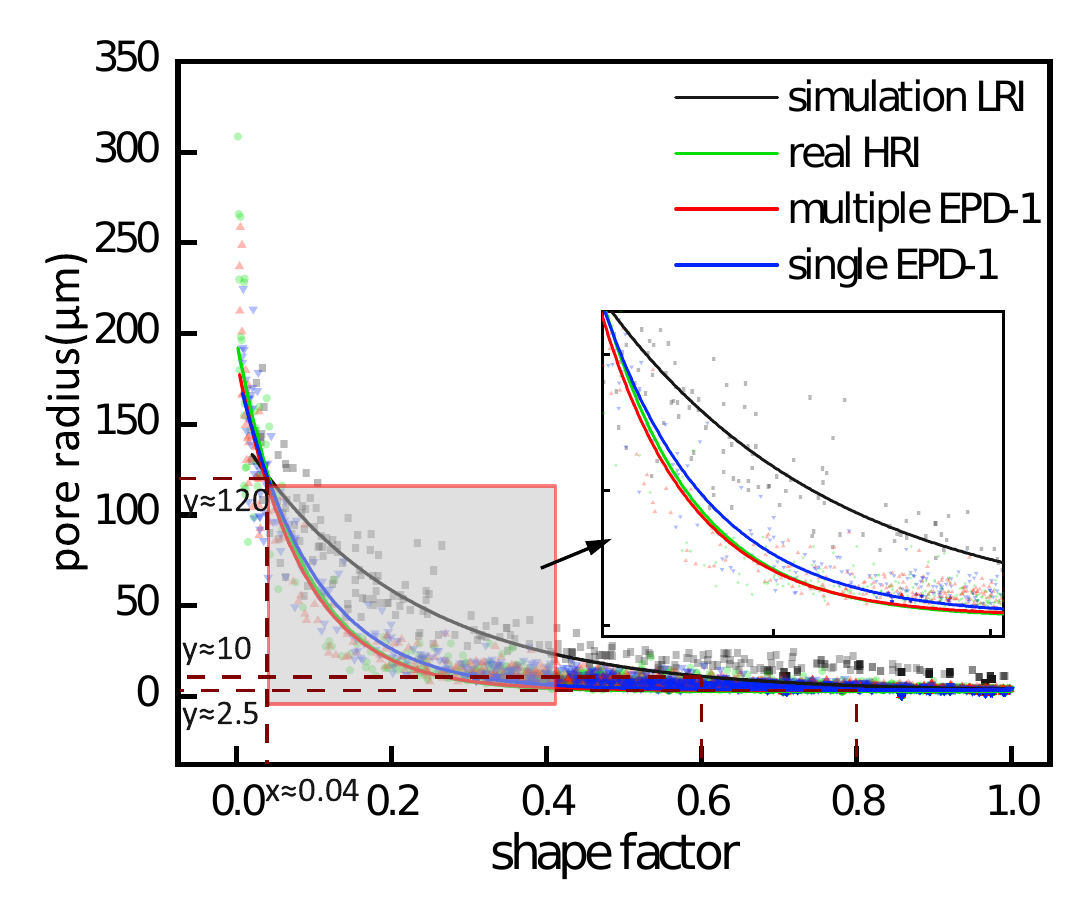}}
	\subfigure[coordination number distribution]{
		\includegraphics[keepaspectratio=True, width=0.48\textwidth]{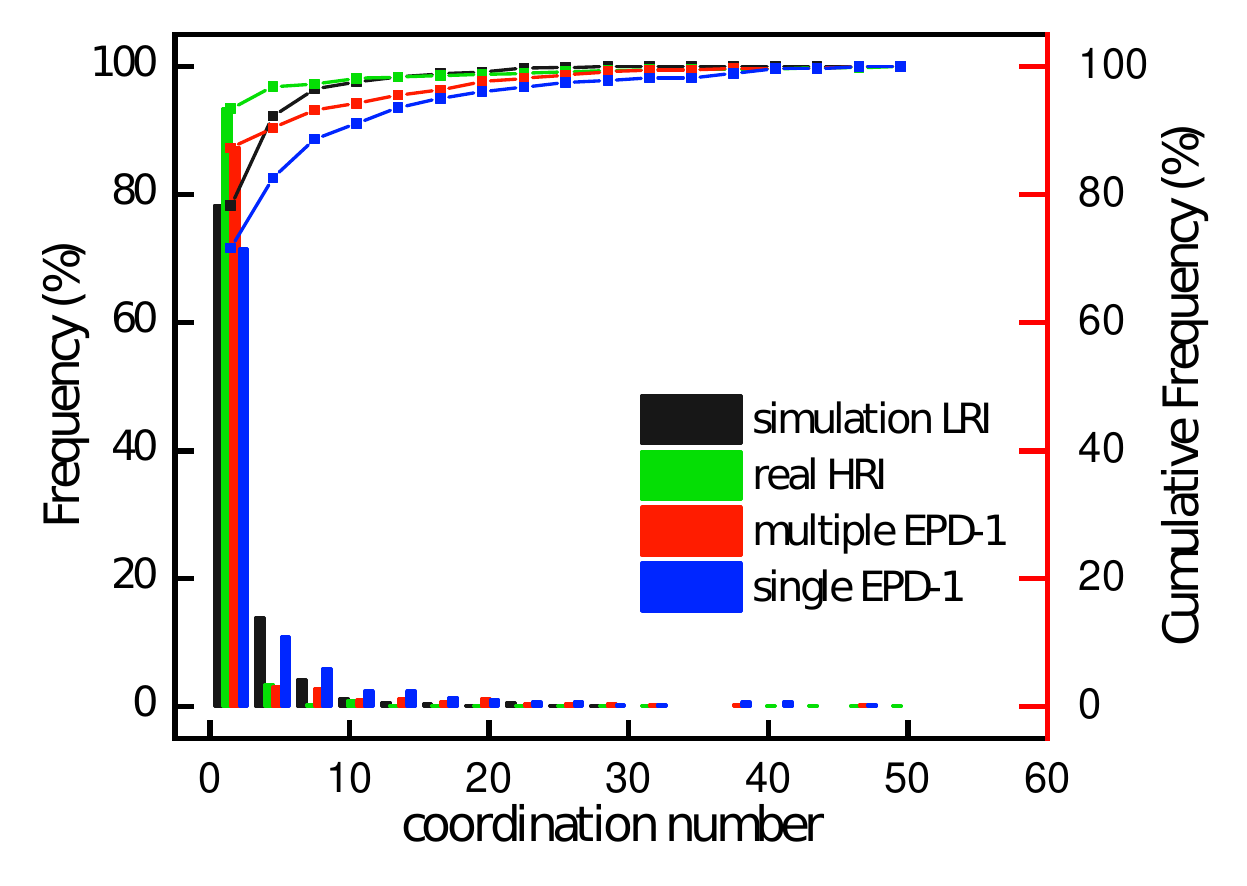}}
	\caption{The comparation of pore radius distribution (a), throat radius distribution (b), shape factor distribution (c) and pore shape factor-pore radius distribution (d), coordination number distribution (e) on sample 1(b) simulation data experiment.}
	\label{fig:Fig12}
\end{figure}

Geometrical and topological features are important physical parameters to evaluate the accuracy of the reconstructions. Figure \ref{fig:Fig12} depicts the comparison of pore radius distribution, throat radius distribution, pore shape factor distribution, pore shape factor-pore radius distribution and coordination number distribution. The Figure \ref{fig:Fig12}(a) shows the count and cumulative frequency of pore radius. Simulation LRI cannot capture micro-pores (here radius smaller than 2.5$\mu \mathrm{m}$), whereas the real HRI and reconstructions all capture micro-pores. It is possible that the difference about the number of micro-pores is caused by the insufficient number of micro-pore patterns in the \emph {MPD}, but the cumulative frequency curve of multiple \emph {EPD}-1 is very similar to that of real HRI. It is worth noting that the same \emph {MPD} is used, but single \emph {EPD}-1 has fewer micro-pores than multiple \emph {EPD}-1. This is because multiple \emph {EPD}-1 introduces pattern information of corresponding scale at each stage of edge reconstruction, which can re-separate some pores adhered to large pores due to low-resolution imaging. From the distribution of throat radius Figure \ref{fig:Fig12}(b) and shape factor Figure \ref{fig:Fig12}(c), the curve trend of multiple \emph {EPD}-1 is more similar to that of real HRI. The fitted curve that counts the shape factor and radius for each pore is the pore shape factor-pore radius distribution, reflecting the trend of the pore shape factor with the pore radius in the pore structure. The shape factor reflects the geometric feature of the pore shape and edges, with the smaller the shape factors the more complex the shape and edges of the pore. From Figure \ref{fig:Fig12}(d) the comparison of shape factor-pore radius distribution, compared with simulation LRI, real HRI and reconstructions have smaller shape factors for pores with the same radius, indicating that the shapes and edges of high-resolution pore structures are more complex. Particularly, the curve of multiple \emph {EPD}-1 is most consistent with that of the real HRI, which demonstrates the effectiveness of multi-stage edge reconstruction by progressively introducing corresponding scale pattern. From Figure \ref{fig:Fig12}(e) the comparison of coordination number distribution, in real HRI, about 98.38\% of the pores has coordination numbers distributed within [0,12], compares to 94.29\% and 91.13\% for multiple \emph {EPD}-1 and single \textit{EPD}-1, respectively. The distribution of coordination number for the multiple \emph {EPD}-1 is more similar to the real HRI. The results in Figure \ref{fig:Fig12} show that reconstructions using multiple edge pattern dictionaries of different scales are in better agreement with the geometric and topological of real samples, and the proposal algorithm is an efficient multiscale reconstruction method. 

\begin{table}[htbp]
	\centering
	\caption{The comparison of the porosity, average pore radius, average throat radius, average coordination number and permeability on sample 1(b) simulation data experiment.}
	\label{table:Table1}
	\begin{tabular}{llllllll}
		\hline  & Simulation & Real & Multiple & \textit{Error $(\%)$ with} & Single & \textit{Error $(\%)$ with} \\
		& LRI & HRI & \textit{EPD}-1 & \textit{Real HRI} & \textit{EPD}-1 & \textit{Real HRI} \\
		\hline Average pore radius/ $\mu \mathrm{m}$ & $9.72$ & $3.96$ & $\mathbf{4.68}$ & $\underline{\textit{18.18}}$ & $6.58$ & $\textit{66.16}$ \\
		Average throat radius $/ \mu \mathrm{m}$ & $17.83$ & $16.46$ & $\mathbf{17.26}$ & $\underline{\textit{4.96}}$ & $18.47$ & $\textit{12.21}$ \\
		Average coordination number & $3.34$ & $2.25$ & $\mathbf{2.76}$ & $\underline{\textit{22.67}}$ & $2.94$ & $\textit{30.67}$ \\
		Permeability $/ \times 10^{-13} \mathrm{m}^{2}$ & $3.9924$ & $3.4471$ & $\mathbf{3.5224}$ & $\underline{\textit{2.18}}$ & $3.6890$ & $\textit{7.02}$ \\
		\hline
	\end{tabular}
\end{table}

The permeability of the core is influenced by a variety of properties such as connectivity, pore shape and pore surface shape. From Table\ref{table:Table1}, the permeability of simulation LRI is $3.9924 \times 10^{-13} \mu \mathrm{m}^{2}$ and the real HRI is $3.4471 \times 10^{-13} \mu \mathrm{m}^{2}$. High-resolution pore structures (real HRI) contain richer details (fine edges and micro-pores) than low-resolution structures (simulation LRI). The pores and throats of the real HRI are finely divided, resulting in smaller average pore radius, average throat radius, and average coordination numbers. Therefore, compared to the simulation LRI, the connectivity of real HRI is less well, while the pore geometry and pore surface morphology are more complex, resulting in lower permeability. By introducing edge patterns and micropore patterns that from other volumes in same core sample, the simulation LRI is reconstructed as a fine multiscale pore structure (multiple \emph {EPD}-1) similar to that of real HRI. This is reflected again in the results in Table 1, the average pore radius, average throat radius and average coordination number of multiple \emph {EPD}-1 are close to that of real HRI. By multiscale reconstruction, the permeability of multiple \emph {EPD}-1  is $3.5224 \times 10^{-13} \mu \mathrm{m}^{2}$, with an error of only 2.18\% from the real HRI. Through the above experiments and analysis, the effectiveness of the proposal algorithm is proved.

\subsubsection{Simulation data experiment for sandstone Sample 2(c)}

\begin{figure*}[htbp]
	\centering
	\includegraphics[keepaspectratio=true, width=0.45\textwidth]{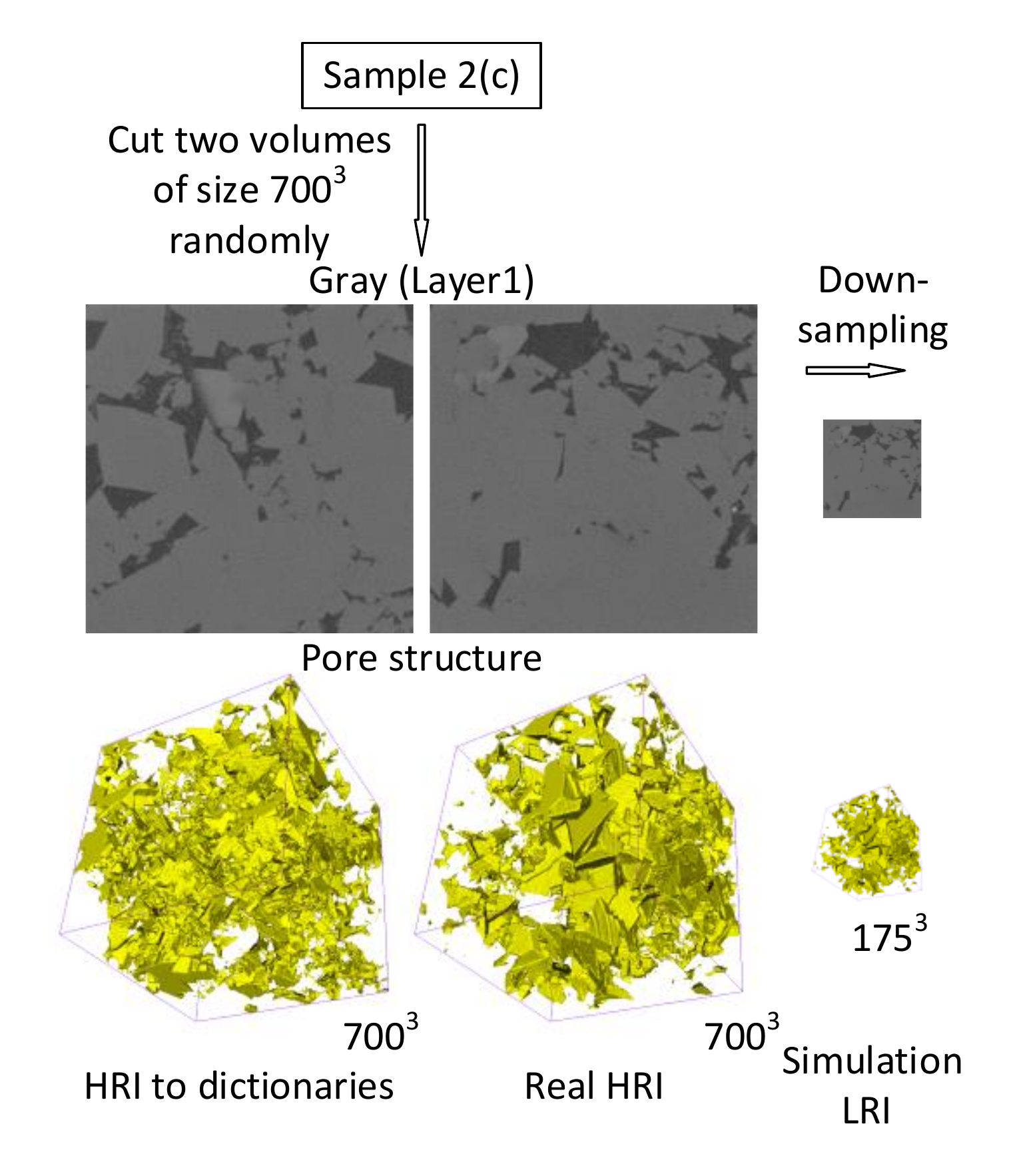}
	\caption{Simulation data for sample 2(c).} 
	\label{fig:Fig13}
\end{figure*}

The proposal method focuses on multiscale reconstruction of large-FoV low-resolution structures using pattern features from small-FoV high-resolution structures. So, this method can realize multiscale reconstruction without the limitation of the size of the low-resolution input pore structure. Simulation data for sample 2(c) are produced following the same processing in section \ref{section:3.1}. Here, the length-scale of the HRI is 1$\mu \mathrm{m}$ and the simulation LRI is 4$\mu \mathrm{m}$. The size of HRI is $700^3$ and the simulation LRI is $175^3$, as shown in Figure \ref{fig:Fig13}. According formula (8), the simulation LRI should be refined edges by a 2-stage edge reconstruction.

\begin{figure*}[htbp]
	\centering
	\includegraphics[keepaspectratio=true, width=0.9\textwidth]{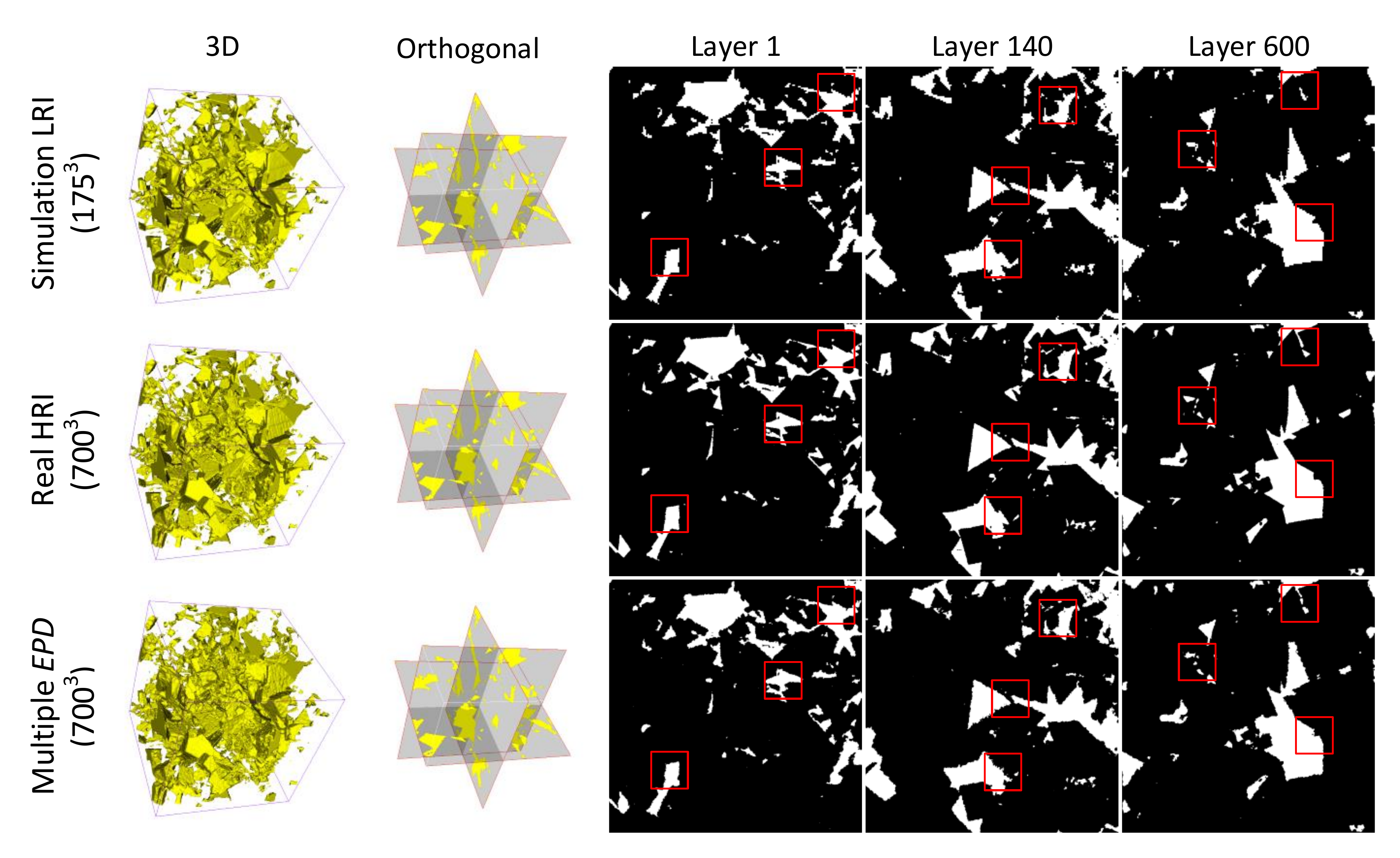}
	\caption{The qualitative comparison of the reconstructions with the real HRI on sample 2(c) simulation data experiment.} 
	\label{fig:Fig14}
\end{figure*}

Figure \ref{fig:Fig14} shows the qualitative comparison of reconstructions with the real HRI. Visually, sample 2(c) is heterogeneous. By multiscale reconstruction, multiple \emph {EPD} acquired edge features similar to real HRI. Even for larger size images ($700^3$), the multiscale reconstructed pore structure of multiple \emph {EPD} is almost agreement to that of real HRI. 

\begin{figure}[htbp]
	\centering
	\subfigure[pore radius distribution]{
		\includegraphics[keepaspectratio=True, width=0.42\textwidth]{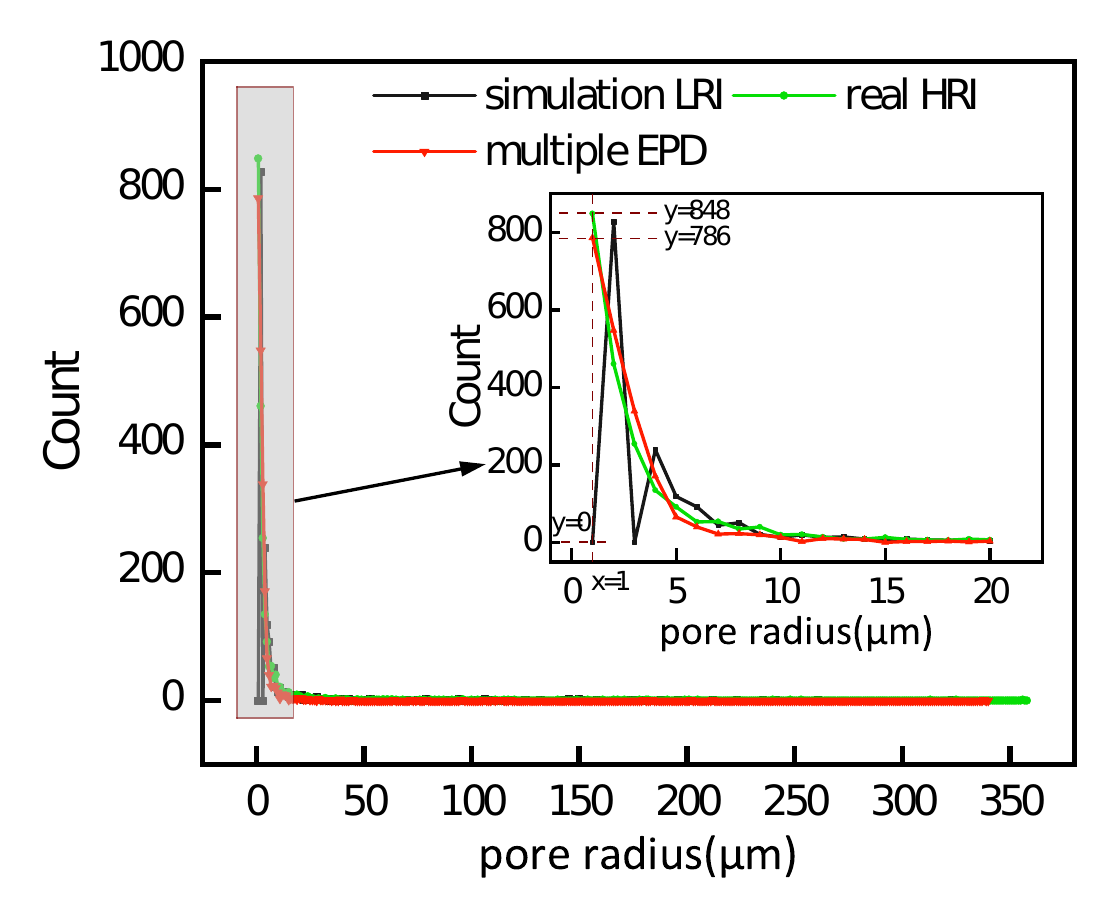}}
	\subfigure[throat radius distribution]{
		\includegraphics[keepaspectratio=True, width=0.40\textwidth]{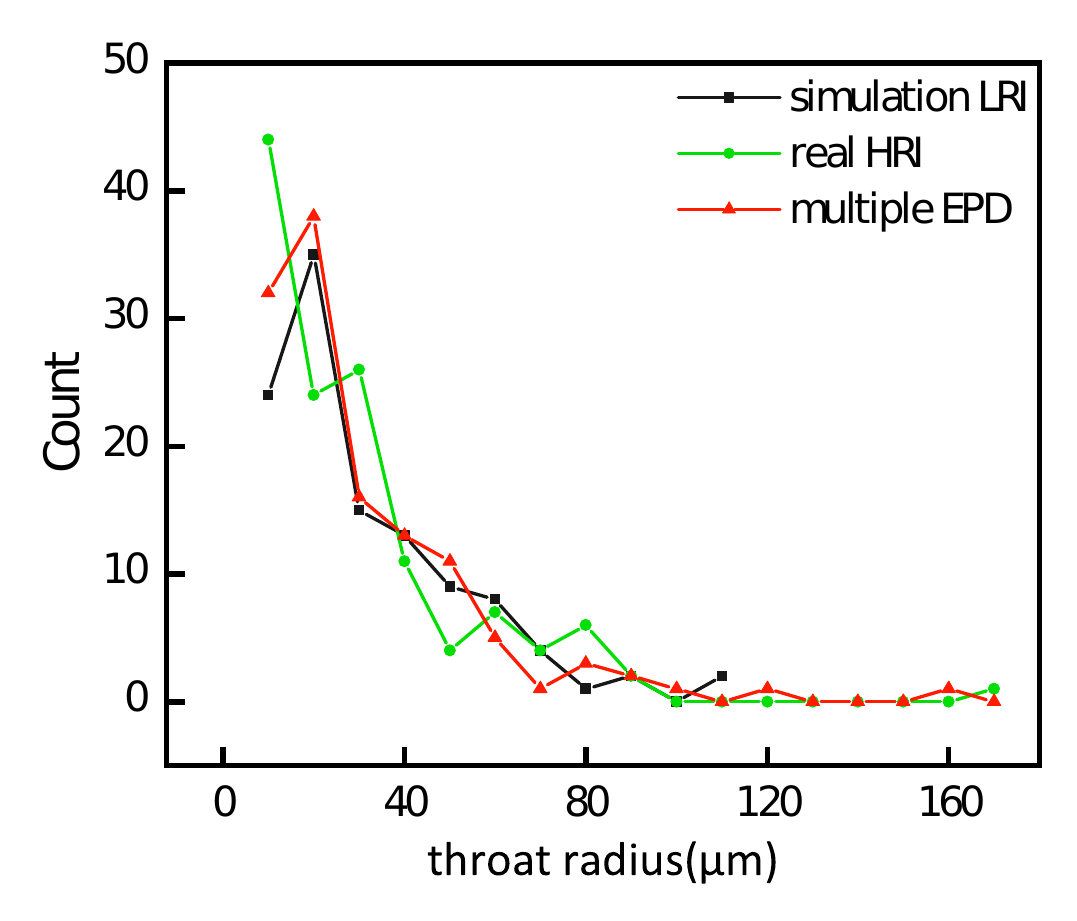}}
	\subfigure[shape factor distribution]{
		\includegraphics[keepaspectratio=True, width=0.42\textwidth]{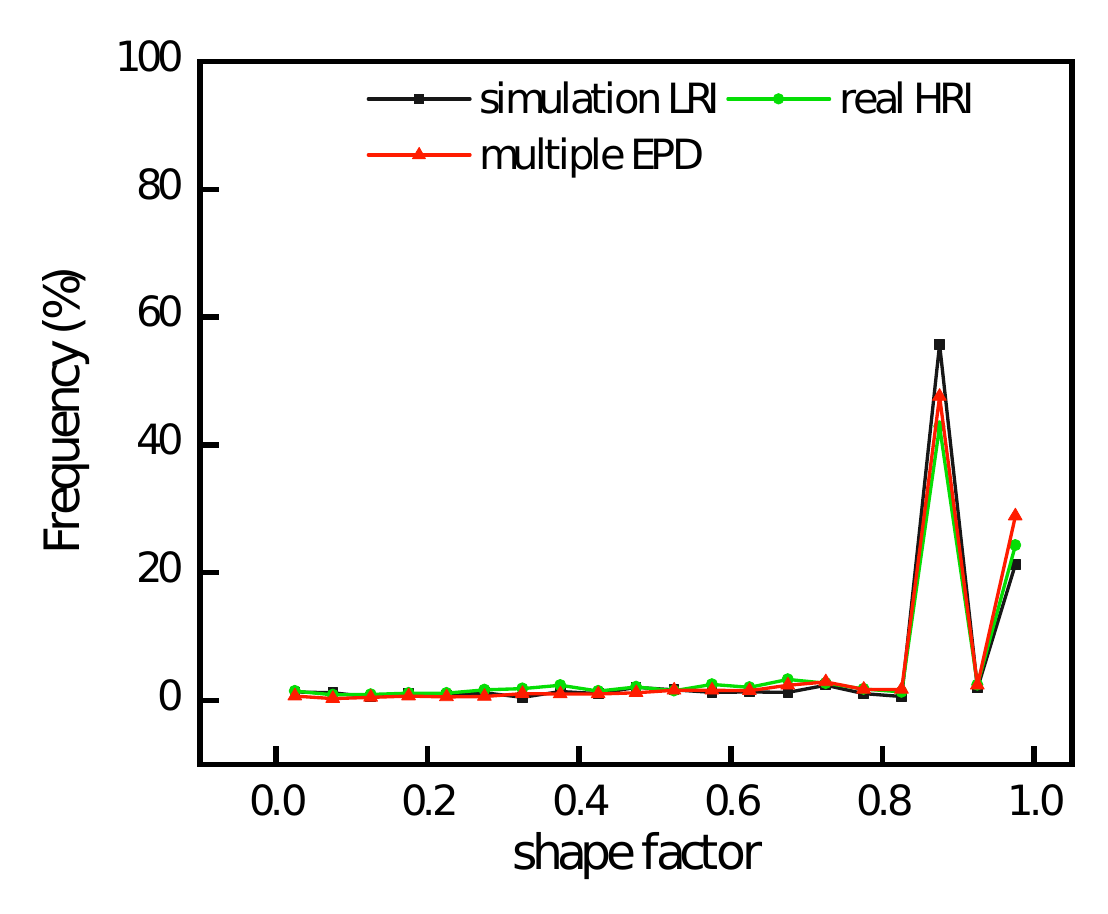}}
	\subfigure[pore shape factor-pore radius distribution]{
		\includegraphics[keepaspectratio=True, width=0.41\textwidth]{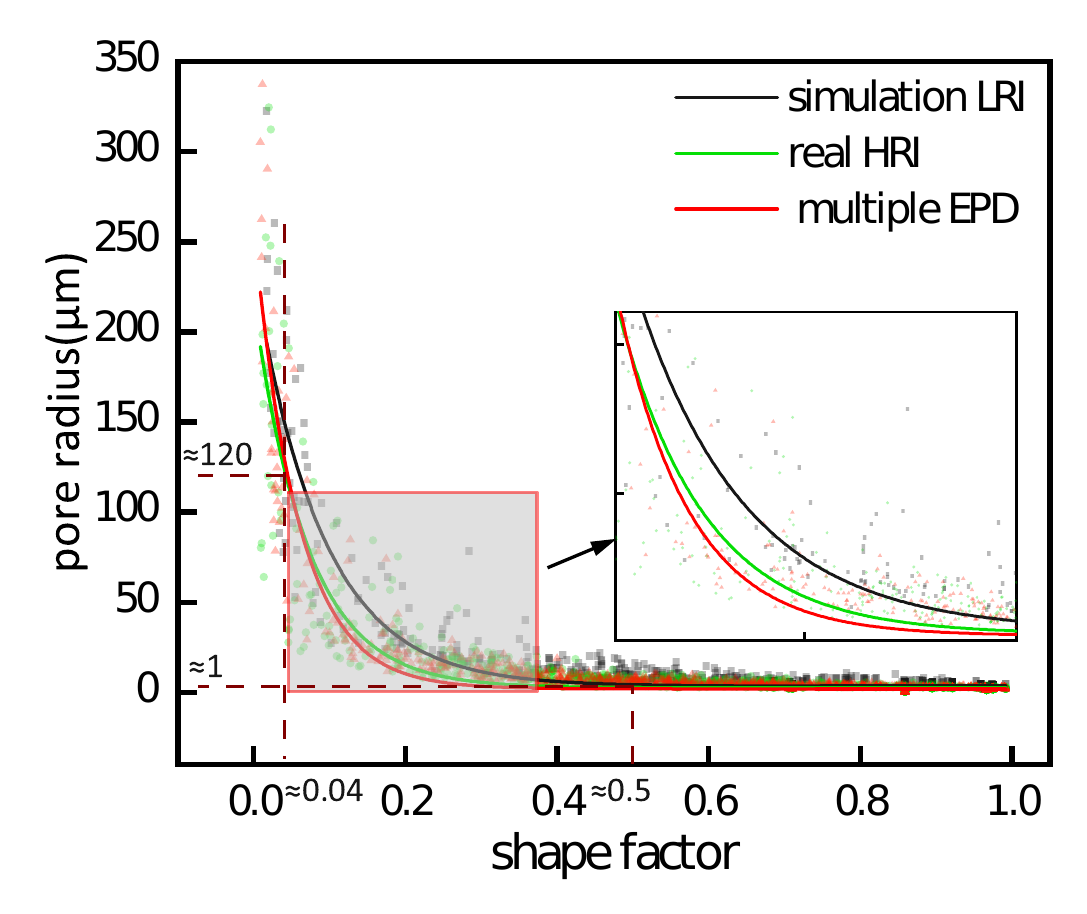}}
	\subfigure[coordination number distribution]{
		\includegraphics[keepaspectratio=True, width=0.48\textwidth]{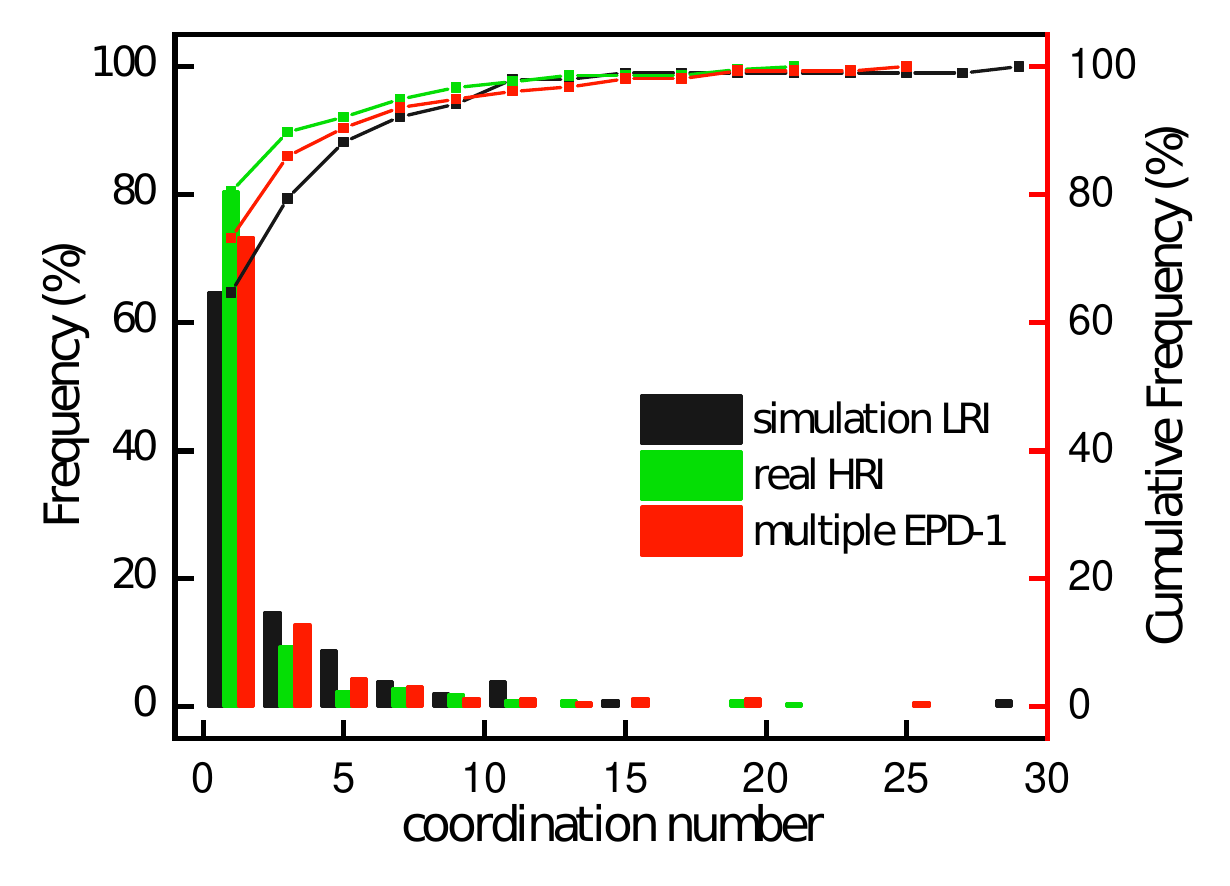}}
	\caption{The comparation of pore radius distribution (a), throat radius distribution (b), shape factor distribution (c) and pore shape factor-pore radius distribution (d), coordination number distribution (e) on sample 2(c) simulation data experiment.}
	\label{fig:Fig15}
\end{figure}

A comparison of the geometric and topological features in Figure \ref{fig:Fig15} further demonstrates the effectiveness of the proposal algorithm. As shown in Figure \ref{fig:Fig15}(a), the real HRI and multiple \emph {EPD} all capture micro-pores (here radius smaller than 1$\mu \mathrm{m}$) that cannot be captured in simulation LRI. From the distribution of throat radius Figure \ref{fig:Fig15}(b), shape factor Figure \ref{fig:Fig15}(c) and shape factor-pore radius Figure \ref{fig:Fig15}(d), all curves trend of multiple \emph {EPD} are similar to that of real HRI. The coordination number distribution of multiple \emph {EPD} is also similar to the real HRI, Figure \ref{fig:Fig15}(e). In real HRI, about 92.09\% of the pores has coordination numbers distributed within [0,6], compares to 90.45\% for multiple \emph {EPD}. The results in Figure \ref{fig:Fig15} show that the proposal algorithm is also effective for multiscale reconstruction of heterogeneous cores. Moreover, experiments based on two sets of different sizes images have also demonstrated that the algorithm is capable of multiscale reconstruction regardless of the input size.

\begin{table}[htbp]
	\centering
	\caption{The comparison of the porosity, average pore radius, average throat radius, average coordination number and permeability on sample 2(c) simulation data experiment.}
	\label{table:Table2}
	\begin{tabular}{llllll}
		\hline  & Simulation & Real & Multiple & \textit{Error $(\%)$ with} \\
		& LRI & HRI & \textit{EPD} & \textit{Real HRI}  \\
		\hline Average pore radius/ $\mu \mathrm{m}$ & $7.97$ & $6.22$ & $\mathbf{6.64}$ & $\underline{\textit{6.75}}$  \\
		Average throat radius $/ \mu \mathrm{m}$ & $26.48$ & $24.94$ & $\mathbf{24.90}$ & $\underline{\textit{0.16}}$  \\
		Average coordination number & $2.65$ & $1.97$ & $\mathbf{2.13}$ & $\underline{\textit{8.127}}$  \\
		Permeability $/ \times 10^{-13} \mathrm{m}^{2}$ & $0.3482$ & $0.3278$ & $\mathbf{0.3355}$ & $\underline{\textit{2.35}}$ \\
		\hline
	\end{tabular}
\end{table}

Although the physical parameters of heterogeneous cores such as permeability are mainly determined by macropores. But the skeleton of the high-resolution pore structure is similar to that of the low-resolution pore structure, and the high-resolution pore structure has a finer delineation of edges and pore throats, so it can be used to calculate more accurate physical parameters. By multiscale reconstruction, fine multiscale pore structures similar to real HRI structures can be obtained. From Table\ref{table:Table2}, the average pore radius, average throat radius and average coordination number of multiple \emph {EPD} are close to that of real HRI. The permeability of multiple \emph {EPD} is $0.3355 \times 10^{-13} \mu \mathrm{m}^{2}$ , with an error of only 2.35\% from the real HRI ($0.3278 \times 10^{-13} \mu \mathrm{m}^{2}$). The above results demonstrate the effectiveness of the proposal algorithm.

\subsection{Real data experiment}

\begin{figure*}[htbp]
	\centering
	\includegraphics[keepaspectratio=true, width=0.45\textwidth]{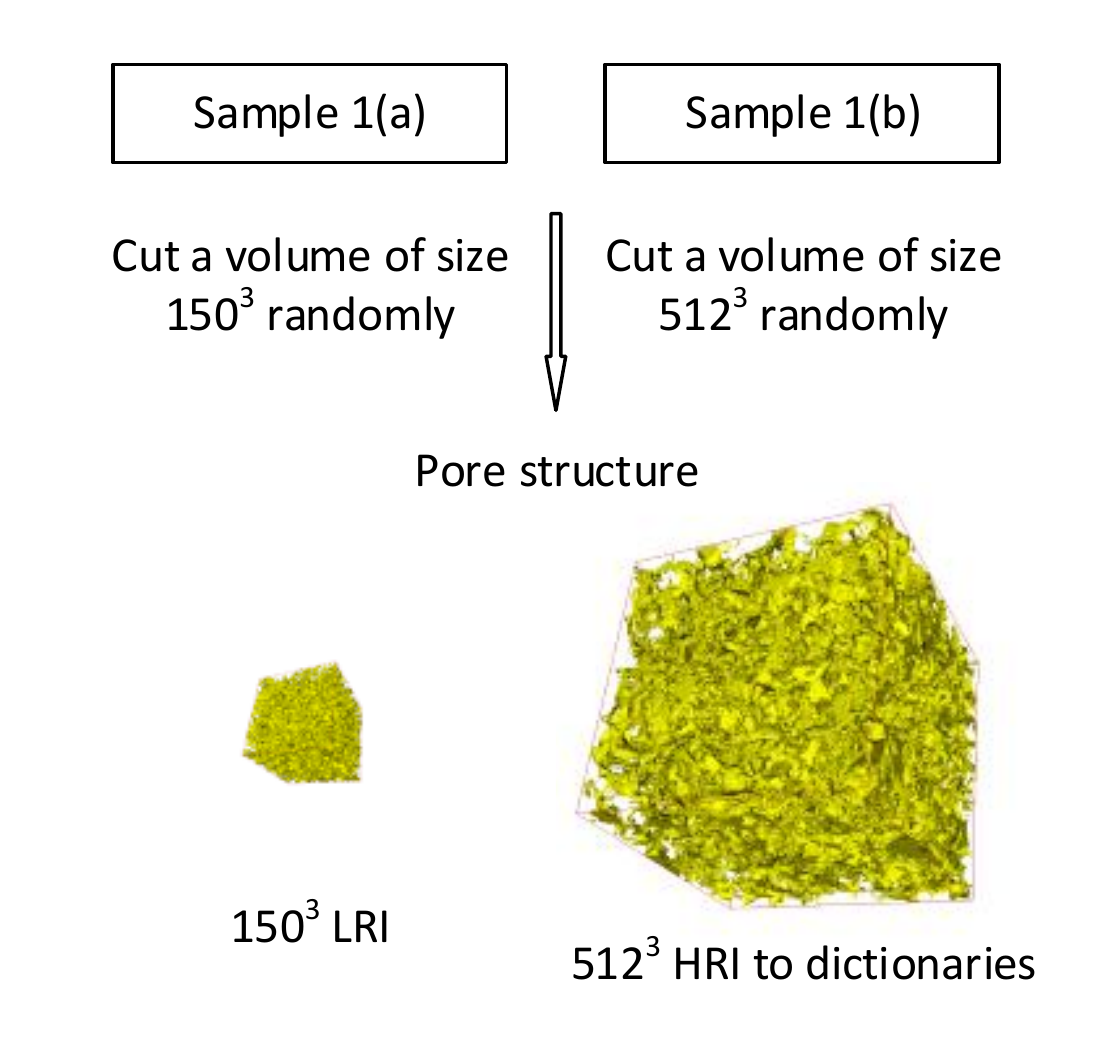}
	\caption{Real data experiment for sample 1(a) and sample 1(b).} 
	\label{fig:Fig16}
\end{figure*}

The real data experiment is conducted to verify the practicality of the proposal algorithm. Sample 1(a) and sample 1(b) are low-resolution images and high-resolution CT images of the same core sample, respectively. The data processing is shown in Figure\ref{fig:Fig16}. Here, the length-scale of the HRI is 2.35$\mu \mathrm{m}$ and the LRI is 13.29$\mu \mathrm{m}$. Multiple dictionaries (include multiple scale edge dictionaries and micropore dictionary) are built by HRI. According formula (4) and formula (8), the LRI is refined edges by a 2-stage edge reconstruction, and the resolution of the reconstructed multiscale pore structure is approximately 3.32$\mu \mathrm{m}$.

\begin{table}[htbp]
	\renewcommand{\arraybackslash}{0}
	\centering
	
	\caption{The comparison of the porosity, average pore radius, average throat radius, average coordination number and permeability on real data experiment.}
	\label{table:Table3}
	\begin{tabular}{llll}
		\hline  & LRI & Multiple \textit{EPD} \\
		\hline Average pore radius/ $\mu \mathrm{m}$ & $46.63$ & $\mathbf{7.42}$  \\
		Average throat radius $/ \mu \mathrm{m}$ & $39.14$ & $\mathbf{36.18}$ \\
		Average coordination number & $4.95$ & $\mathbf{2.74}$   \\
		Permeability $/ \times 10^{-13} \mathrm{m}^{2}$ & $52.7318$ & $\mathbf{46.2224}$  \\
		\hline
	\end{tabular}
\end{table}

\begin{figure*}[htbp]
	\centering
	\includegraphics[keepaspectratio=true, width=1\textwidth]{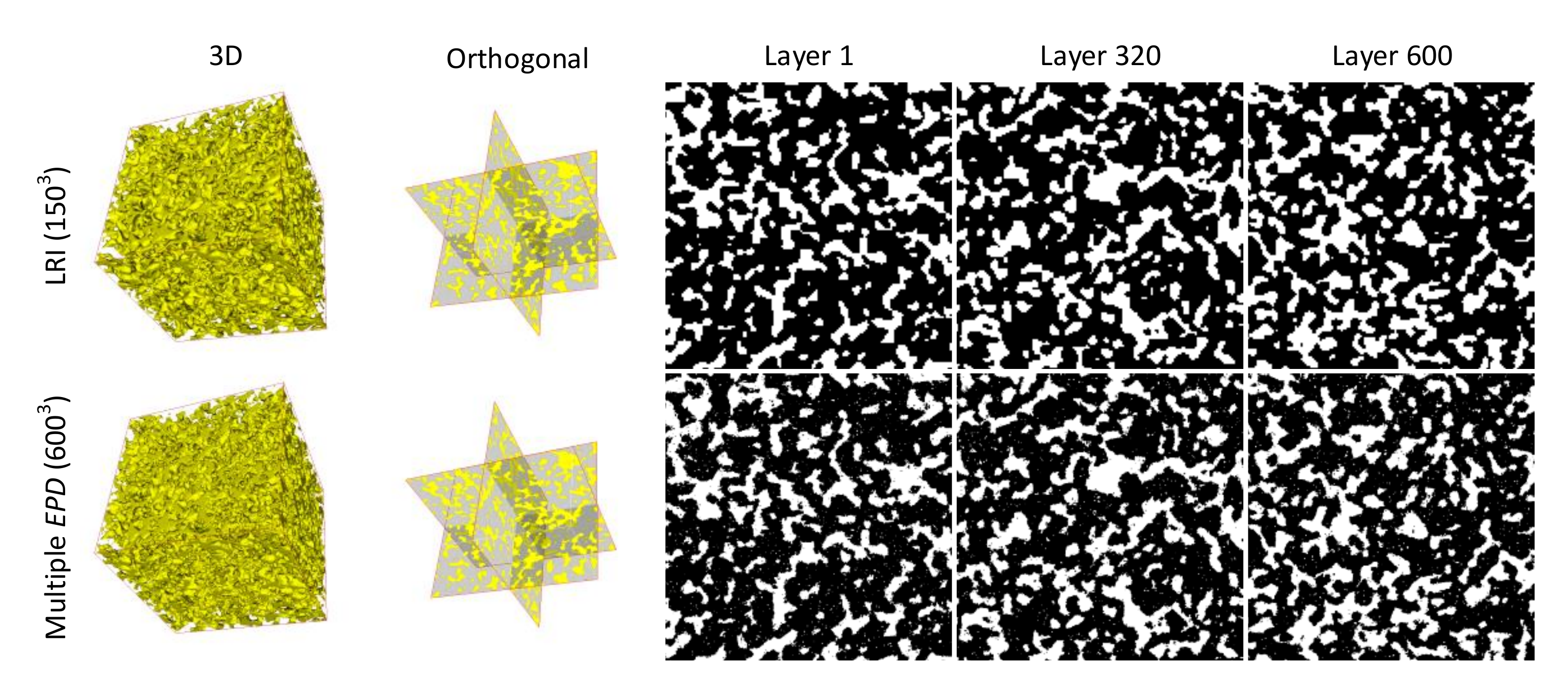}
	\caption{The qualitative comparison of the reconstructions with the real HRI on real data experiment.} 
	\label{fig:Fig17}
\end{figure*}

Visually, the multiple \emph {EPD} has finer pore edges than the LRI, while capturing the micro-pores (here radius smaller than 4$\mu \mathrm{m}$), in Figure\ref{fig:Fig17}. From Table 3, the average pore radius and average throat radius, and the average coordination number have decreased. This indicates that the pores and throats of multiple \emph {EPD} are finely divided, because the multiscale reconstruction introduces rich edge information and micropore information. Moreover, the permeability is reduced from (LRI) $52.7318 \times 10^{-13} \mu \mathrm{m}^{2}$ to (multiple \emph {EPD}) $46.2224 \times 10^{-13} \mu \mathrm{m}^{2}$, which is consistent with the conclusions of the simulation data experiments. The effectiveness and practicality of the proposal algorithm is demonstrated by real data experiment.

\section{\label{sec:conclusion}Conclusion}

In this paper, a multiple dictionaries learning-based multiscale reconstruction method is proposed for reconstructing large-FoV high-resolution pore structures using large-FoV low-resolution pore structures and small-FoV high-resolution pore structures. Based on the homology and statistical similarity, multiple dictionaries are built by extracting the multiscale edge patterns and micro-pores pattern in HRI, and then reconstruct pore edges of LRI by matching the corresponding scales edge patterns stage by stage. Meanwhile, masks are adopted to avoid the conglomeration of micro-pore patterns to pore areas. Through experiments on simulation data and real data, the effectiveness and practicality of this algorithm are proved from quantitative and qualitative comparison. Experiments show that multiscale reconstruction by introducing high-resolution edge patterns and micro-pore patterns from different volumes of the same core, allows the results of multiscale reconstruction to obtain physical properties similar to those of high-resolution core structures. The algorithm is capable of multiscale reconstruction without regard to the size of the input. Moreover, it is all effective for multiscale reconstruction of homogeneity and heterogeneity core. In this paper, the random padding of the micro-pores is primarily based on statistical quantity similarities in homology cores, but the location of the micro-pores also affects the accuracy of the multiscale reconstruction, which will be explored in a subsequent study.

\section*{Acknowledgment} 

This work was supported by the National Natural Science Foundation of China (Grants No. 62071315). 

\section*{References}

\bibliography{References}

\end{document}